\documentclass[fleqn,useAMS,usenatbib]{mnras}

\usepackage{newtxtext,newtxmath}
\usepackage[T1]{fontenc}
\usepackage[dvipsnames]{xcolor}
\usepackage{wasysym}
\usepackage{graphicx}
\usepackage{hyperref}  
\usepackage{xspace}
\usepackage{caption}
\usepackage{placeins}
\usepackage{xcolor}

\newcommand{\msun}{\,M$_{\odot}$\xspace}

\volume{{\rm submitted}}

\title{Ionizing photon production and escape fractions during cosmic reionization in the TNG50 simulation}
\author[I. Kostyuk et al.]{Ivan Kostyuk$^{1,2}$\thanks{E-mail: ivkos@mpa-garching.mpg.de}, Dylan Nelson$^3$, Benedetta Ciardi$^1$, Martin Glatzle$^1$, Annalisa Pillepich$^4$
\\\\
$^{1}$Max-Planck-Institut f\"{u}r Astrophysik, Karl-Schwarzschild-Str. 1, 85741 Garching, Germany\\
$^{2}$Ludwig-Maximilians-Universit{\"a}t M{\"u}nchen (LMU), Geschwister-Scholl-Platz 1, 80539 M{\"u}nchen, Germany\\
$^{3}$Universit\"{a}t Heidelberg, Zentrum f\"{u}r Astronomie, Institut f\"{u}r theoretische Astrophysik, Albert-Ueberle-Str. 2, 69120 Heidelberg, Germany\\
$^{4}${Max-Planck-Institut f{\"u}r Astronomie, K{\"o}nigstuhl 17, 69117 Heidelberg, Germany} \\
}

\begin{document}
\maketitle

\begin{abstract}
In this work we investigate the dependence of the escape fraction of ionizing photons, $f_{\rm esc}$, on various galaxy and host halo properties during the epoch of reionization. We post-process the TNG50 magneto-hydrodynamical simulation from the IllustrisTNG project using the 3D multi-frequency radiative transfer code CRASH. Our work covers the stellar mass range $10^6 \lesssim M_\star/{\rm M_\odot} \lesssim 10^8$ at redshifts $6 < z < 10$. Adopting an unresolved, cloud-scale escape fraction parameter of unity, the halo escape fraction $f_{\rm esc}$ increases with mass from $\sim 0.3$ at $M_\star = 10^6$\,M$_\odot$ to $\sim 0.6$ at $M_\star = 10^{7.5}$\,M$_\odot$, after which we find hints of a turnover and decreasing escape fractions for even more massive galaxies. However, we demonstrate a strong and non-linear dependence of $f_{\rm esc}$ on the adopted sub-grid escape fraction. In addition, $f_{\rm esc}$ has significant scatter at fixed mass, driven by diversity in the ionizing photon rate together with a complex relationship between (stellar) source positions and the underling density distribution. The global emissivity is consistent with observations for reasonable cloud-scale absorption values, and halos with a stellar mass $\lesssim 10^{7.5}$\,M$_\odot$ contribute the majority of ionizing photons at all redshifts. Incorporating dust reduces $f_{\rm esc}$ by a few percent at $M_\star \lesssim 10^{6.5}$\,M$_\odot$, and up to 10\% for larger halos. Our multi-frequency approach shows that $f_{\rm esc}$ depends on photon energy, and is reduced substantially at $E>54.4$eV versus lower energies. This suggests that the impact of high energy photons from binary stars is reduced when accounting for an energy dependent escape fraction.
\end{abstract}



\section{Introduction}

The last major phase transition in the Universe, the reionization of the intergalactic medium (IGM), took place during its first billion years. During this era matter collapsed into bound objects and the first stars and galaxies formed, at redshifts as high as $z \approx 30$ \citep{loeb01}. The radiation produced by these first luminous sources was able to eventually ionize the gas of the diffuse IGM.

The Planck measurements of the Thomson scattering optical depth, of CMB photons scattered by the free electrons released during reionization, give a value of $\tau_e = 0.056 \pm 0.007$ \citep{planck18}. Given the scenario of an instantaneous reionization process, this corresponds to a redshift $z_{\rm ion} = 7.68 \pm 0.79$ for complete reionization. However, the reionization history of the Universe is much more complex and depends, at a minimum, on the detailed distribution, timing, and properties of the sources of ionizing radiation.

Both active galactive nuclei (AGN; e.g. \citealt{madau15}) and star forming galaxies (e.g. \citealt{robertson13}) have been suggested as the main drivers of reionization. However, recent semi-analytic (e.g. \citealt{qin17}) and semi-numeric (e.g. \citealt{hassan18}) models, hydrodynamic plus radiation transfer (RT) simulations (e.g. \citealt{eide20}), as well as measurements of AGN at $z \approx 6$ \citep{aloisio17, parsa18} indicate that the abundance of these sources at high $z$ is likely too low to dominate the ionizing photon budget. This leaves galaxies as the most likely drivers of reionization. Nevertheless, it is not well understood which galaxies contribute the most, and how the diversity of physical properties of galaxies control their relative efficiency of ionizing photon emission.

So far, the reionization process has been modelled with different methods. Semi-analytic models (e.g. \citealt{barkana00, furlanetto04, choudhury07,raicevic11,qin17}) have been generally applied to describe the evolution of the mean properties of the IGM, and are particularly useful when investigating parameter dependences due to their speed and flexibility. In semi-numeric models (e.g. \citealt{santos10, mesinger11, furlanetto16}) the density distribution is derived based on e.g. an excursion set formalism (but see e.g. also the N-body simulation based \citealt{Hutter2021}), and used in combination with assumptions on source properties and the formation of ionized bubbles, in order to model the reionization process. 

Finally, there are a number of approaches that combine N-body/hydrodynamical simulations and RT calculations. In this context a trade off is made between large scales, needed to properly capture the evolution of large ionized regions -- which, towards the end of the reionization process, can measure tens of cMpc (see e.g. \citealt{busch20}) -- and small scales, needed to model the properties of the stars and interstellar medium gas within galaxies themselves. The inclusion of RT adds further computational complexity and can be treated either in post-processing or directly on-the-fly with coupled radiation-hydrodynamical (RHD) simulations \citep[see also 1D approaches;][]{thomas2009,ghara2015}. Post-processing enables large scale ($\gtrsim 100 h^{-1}$~cMpc) simulations, and the inclusion of more expensive, multi-frequency treatments \citep{Iliev2014,eide20}. On the other hand, smaller scale simulations in which the RT equation is solved together with the hydrodynamics offer a self-consistent coupling with the underlying gas dynamics and astrophysical models \citep{gnedin2000,baek2010,gnedin14,pawlik17,rosdahl18,ocvirk21,kannan22}.

A crucial quantity controlling the reionization process is the `escape fraction' $f_\mathrm{esc}$, the ratio of ionizing photons which escape into the IGM relative to the total produced. The escape fraction is sensitive to the small scale production and propagation of ionizing Lyman continuum (LyC) photons in the galaxy. Simultaneously, the escape fraction also controls the large scale ionization of the IGM. 

Observationally, it is not currently possible to directly identify high redshift leakers of LyC photons, and measurements of $f_\mathrm{esc}$ can only be performed at lower redshifts with relatively small samples of galaxies (see e.g. \citealt{japelj17, bian20, choi20, mestic21, naidu21}). Therefore, it has been suggested that other properties of LyC emitters may correlate with a high transmission of ionizing photons, and that these observables could be used as proxies for $f_\mathrm{esc}$. Possible properties of this type include the strength and details of the Lyman-alpha line of hydrogen (\citealt{dijkstra14, vanzella17, chisholm18}), the [O III]/[O II] optical line ratio (\citealt{faisst16, chisholm18, katz20, barrow20}), as well as emission from the Mg II doublet at 2796\AA~ and 2803\AA~ \citep{xu22}. However, these proxies are not perfect: the ratio between the Ly$\alpha$ and ionizing luminosity is not precisely determined \citep{inoue05, guaita16}, a large [O III]/[O II] ratio is only a necessary but not sufficient condition for strong LyC leakage (see \citealt{nakajima20}), and the Mg II doublet emission tends to correlate with the dust content rather than with neutral hydrogen, limiting its use to only finding LyC leakers in the optically thin regime \citep{katz22}. Additionally, $f_\mathrm{esc}$ values measured at low redshift are not directly applicable to higher redshifts, as the physical properties of galaxies and their source populations evolve (\citealt{yajima09, razoumov10, ferrara13, dijkstra14, wise14, sharma16, price16}).

It is also possible to directly estimate the escape fraction of ionizing radiation from numerical simulations, and numerous cosmological RHD simulations have been used for this purpose. These include the CoDa \citep{ocvirk16} and CoDa-II \citep{ocvirk21} simulations, which use the RAMSES code and are run in a 95cMpc box with a dark matter (DM) particle mass of $m_\mathrm{dm}=4.07 \times 10^5$M$_{\astrosun}$. Other RHD simulations employing various physical models within RAMSES include those by \citet{kimm14} and \citet{ocvirk21}, which have box sizes of 25$h^{-1}$cMpc and 8$h^{-1}$cMpc, respectively, with a particle mass of $m_\mathrm{dm}=1.6 \times 10^5$M$_{\astrosun}$ and $m_\mathrm{dm}=5 \times 10^4$M$_{\astrosun}$. The SPHINX simulations have realized boxes with side lengths of $5$, $10$ and $20$cMpc, with a mass resolution of $m_\mathrm{dm}=3.1\times 10^4$M$_{\astrosun}$ and $2.5\times 10^5$M$_{\astrosun}$ respectively \citep{rosdahl18, rosdahl22}. These are slightly smaller volumes, with higher resolution, than the CROC \citep{gnedin14} and Aurora \citep{pawlik17} simulations. Most recently, the THESAN simulation couples the IllustrisTNG galaxy formation model to a RT solver, running a $95.5$cMpc box with mass resolution of $m_\mathrm{dm}=3.1\times 10^6$\,M$_{\astrosun}$. \citep{kannan22,garaldi22,smith22}. 

Broadly speaking, most studies of halos in the mass range $\approx 10^{6}-10^{11}$M$_{\astrosun}$ find that $f_\mathrm{esc}$ decreases with increasing halo mass (e.g. \citealt{ricotti2000,wood2000,kitayama2004,razoumov10,kimm14}) and decreasing redshift (e.g. \citealt{fujita2003,razoumov10,kimm14}). The physics behind the scaling of escape fraction with galaxy mass, and galaxy properties, are complex. The giant molecular clouds in which stars form are dense and therefore effective at absorbing radiation. At the same time, they host most of the short lived O and B stars that produce the bulk of ionizing radiation. For this reason, runaway OB stars which leave their formation sites early and therefore emit LyC radiation in a less opaque environment are thought to be a major contributor to LyC escape (e.g. \citealt{conroy12,kimm14}). Similarly, because binary systems emit ionizing photons on timescales longer than single stars, they continue to produce ionizing radiation also after the dissolution of their birth clouds, providing a major contributing factor to the escape of LyC photons \citep{ma16,rosdahl18,ma20}. Finally, the rapidly expanding super-bubbles produced by young stars (e.g. \citealt{dove2000,ma20}) are also crucial in aiding the escape of LyC photons from such environments.

In order to explore the escape of ionizing radiation through the epoch of reionization, we turn to the TNG50 magneto-hydrodynamic (MHD) simulation \citep{pillepich19,nelson19a}. We post-process halos at $6 < z < 10$ with the Monte Carlo RT code CRASH \citep{crash,crash3,glatzle21}, thereby incorporating the impact of ionizing radiation on the surrounding hydrogen and helium. 

While simulations which directly couple radiation to hydrodynamics are able to capture the radiation-gas back-reaction self-consistently, the Monte Carlo approach we adopt here has complementary strengths. It allows for a more accurate solution of the radiative transfer equation, capturing better effects such as shadowing, as well as a much better sampling of the source spectral energy distributions. Our approach in this respect is similar to that adopted by \cite{paardekooper15} for the First Billion Year project and by \cite{ma20} for the FIRE-II simulations. 

The paper is structured as follows. In Section \ref{sec:methods} we describe the simulations and our analysis methodology. We present our main results, including the inferred escape fraction from galaxies, in Section \ref{sec:fesc}. In Section \ref{sec:conlusions} we summarize our main conclusions, their implications, and discuss future steps.


\section{Method}\label{sec:methods}

\subsection{The IllustrisTNG simulations}

The IllustrisTNG project \citep{pillepich18b,springel18,naiman18,nelson18a,marinacci18} consists of three different cosmological large-scale simulations, TNG300, TNG100 and TNG50 \citep{nelson19a,pillepich19}, covering volumes of $(302.6\mathrm{Mpc})^3$, $(110.7\mathrm{Mpc})^3$ and $(51.7\mathrm{Mpc})^3$, respectively. The simulations were run from a redshift of $z=127$ to the present day, and tested in this stringent low redshift regime. This provides some confidence that the higher redshift halos we post-process are realistic progenitors. The initial conditions were generated using the N-GenIC code \citep{springel05} and cosmological parameters consistent with values from the \citet{planck16}, namely $\Omega_\mathrm{m}=0.3089$, $\Omega_\mathrm{b}=0.0486$, $\Omega_\Lambda=0.6911$, $h=0.6774$, $\sigma_8=0.8159$ and $n_s=0.9667$. 

The TNG simulations were run with the {\textsc AREPO} code \citep{springel10}, which is used to solve the equations of ideal MHD \citep{pakmor11} describing the non-gravitational interactions of baryonic matter, as well as the gravitational interaction of all matter. All use the same TNG galaxy formation model \citep{weinberger17,pillepich18a}. Star formation occurs by stochastically converting gas cells into star particles above a density threshold of $n_\mathrm{H} \approx 0.1\,\mathrm{cm}^{-3}$ following the Kennicutt-Schmidt relation \citep{springel03}. For these stars a Chabrier initial mass function \citep[IMF;][]{chabrier03} is assumed. The TNG simulations track chemical enrichment of the gas, following H, He, C, N, O, Ne, Mg, Si, Fe and Europium, as well as the total gas metallicity. Stellar feedback from supernovae is included with a decoupled kinetic galactic wind model \citep{pillepich18a}. In addition, supermassive black holes (SMBHs) are treated as sink particles which form with a seed mass of $8\times 10^{5}h^{-1}$\msun at the center of halos which surpass a total mass threshold of $10^{10.8}$\msun. Subsequently they grow via gas accretion following the Bondi formalism, limited to the Eddington rate, as well as via SMBH-SMBH mergers. Feedback from SMBHs is modeled via a multi-mode scheme, including thermal heating and radiative influences (at high accretion rates), as well as a kinetic wind (at low accretion rates), all depositing energy into the surrounding gas \citep{weinberger17}.

While the simulations do not include on-the-fly RT and radiation from local ionizing sources, the time-variable though spatially uniform UV-background (UVB) radiation field of \citet{fg09} (FG11 version) is turned on at $z < 6$ \citep{pillepich18a}. Note that this differs from the gradual build up at redshift $6 < z < 9$ as prescribed by \citet{fg09}. Recent updates and many alternative models for the evolving UVB exist \citep{fg20,puchwein19}, with implications for low-density gas heating.

\begin{figure}
    \centering
    \includegraphics[width=0.47\textwidth]{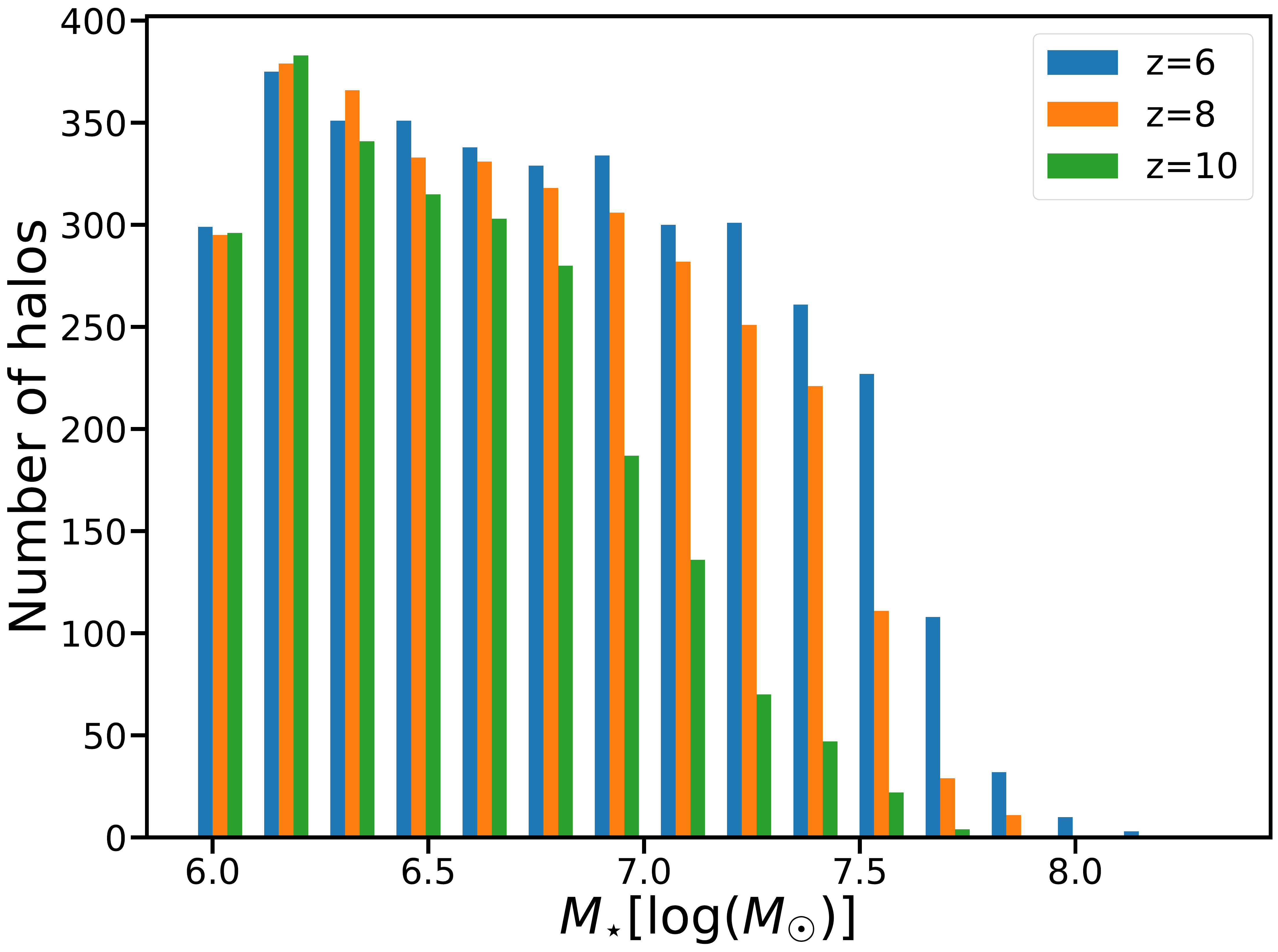}
    \includegraphics[width=0.47\textwidth]{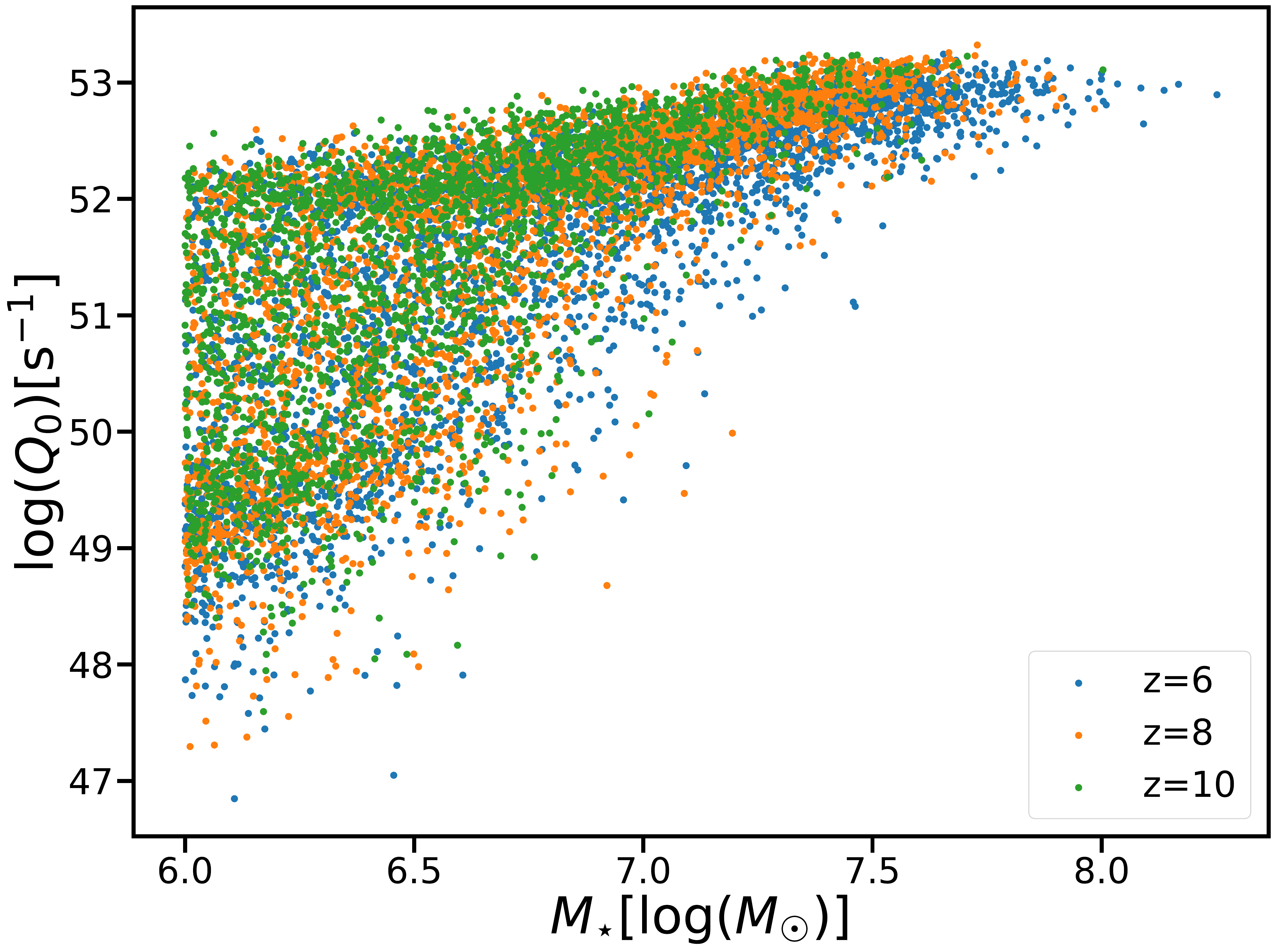}
    \caption{{\it Top panel}: Number of halos post-processed with CRASH RT in a given mass bin for the fiducial configuration (see Table \ref{tab:configs}). {\it Bottom}: Intrinsic emissivity, in ionizing photons per second, of each halo. Our complete sample at redshifts $z=6, 8, 10$ is shown by the blue, orange, and green symbols, respectively. At all redshifts we restrict to a galaxy stellar mass range of $10^{6} < M_{\mathrm{\star}} / \rm{M}_\odot \lesssim 10^{8}$.
    }
    \label{fig:sample}
\end{figure}

In this work we exclusively use the highest resolution simulation, TNG50 \citep{pillepich19,nelson19a}, with a dark matter particle mass resolution of $4.5\times10^5$\msun. The initial mass of gas cells is $8.5\times10^4$\msun, and during the simulation this value evolves within a factor of two due to the refinement and de-refinement of Voronoi cells. Stellar particles inherit the mass of the baryonic gas cell they are formed in, and then lose mass due to stellar evolution processes. For reference, at $z=6$ the star-forming gas cells reach a minimum of 2.7 physical parsecs in size, while the mean (median) size of star-forming gas is 144 pc (154 pc) with 16-84th percentiles from 96 to 184 pc.

Halos are obtained with the {\textsc SubFind} algorithm \citep{springel01}, which runs on friends-of-friends (FoF) identified structures \citep{davis85}. Subhalos and halos require a minimum of 20 and 32 particles, respectively, such that halos with a total mass down to $\sim\,10^{6-7}$\msun are identified. In this work we consider only central galaxies above a minimum stellar mass of $M_\star \geq 10^{6}$\msun to ensure they are reasonably resolved, with at least 20 stellar particles and $\gtrsim 500$ gas cells. In contrast, the most massive galaxies in our sample contain $\sim 20,000$ stars and $\sim 300,000$ gas cells. At $z=6, 8$ and $10$ there are 36,000, 14,000, and 5,000 galaxies sufficiently massive, respectively. From these large samples we select smaller random subsets in bins of log stellar mass for further processing.

The top panel of Figure \ref{fig:sample} shows the number of halos selected for our analysis in each stellar mass bin. Thanks to the large parent volume, our sample provides ample statistics over the range of low-mass galaxies that are thought to be the primary drivers of cosmic hydrogen reionization.

The bottom panel of Figure \ref{fig:sample} depicts the total ionizing emissivity of the halos post processed with the CRASH RT simulation as a function of their stellar mass. We find that the emissivities at a given stellar mass are similar for halos at all redshifts. We also see that the variation in the emissivities decreases with increasing stellar mass. 

\subsection{Radiation transfer with CRASH}

The cosmological radiative transfer scheme for hydrodynamics (CRASH) is a Monte Carlo algorithm for modeling radiation transfer on a Cartesian grid \citep{crash}. It captures the interaction of ionizing radiation with hydrogen and helium gas \citep{crash_upgrade, crash2}, as well as atomic metals \citep{crash3}. Spectral distributions of sources are accounted for through frequency-binned photon packets. Recently, absorption due to dust grains has also been incorporated \citep{dust,glatzle21}.

Each CRASH simulation requires as input the total gas number density, $n_{\mathrm{gas}}$, in each grid cell, as well as the initial physical properties of the gas, i.e. its temperature, $T$, and the fraction, $x_i$, of each considered ionization state $i$. Finally, a list of the position, spectral energy distribution (SED) and emission rate of ionizing photons, $\dot{N}_j$, of all point sources is needed as well\footnote{Note that each quantity should be written also as an explicit function of the grid cells, e.g. $T^c$, but for the sake of clarity we omit such additional index.}.

\begin{table}
 \centering
\begin{tabular}{|c||c|c|c|c|c|c|c|}
\hline
Run & Photon packets & $f_\mathrm{esc, loc}$ & Dust \\
\hline \hline
\textbf{fiducial} & $\mathbf{10^8}$ & $\mathbf{1.0}$ & \textbf{no} \\ 
\hline
dust & $\mathbf{10^8}$ & $\mathbf{1.0}$ &  yes\\
\hline
fesc\_0.1 & $\mathbf{10^8}$ & $0.1$ &  $\mathbf{no}$\\ 
\hline
fesc\_0.3 & $\mathbf{10^8}$ & $0.3$ &  $\mathbf{no}$\\ 
\hline
fesc\_0.5 & $\mathbf{10^8}$ & $0.5$ &  $\mathbf{no}$\\ 
\hline
fesc\_0.7 & $\mathbf{10^8}$ & $0.7$ &  $\mathbf{no}$\\
\hline
conv\_5e6 & $5 \times 10^6$ & $\mathbf{1.0}$ &  $\mathbf{no}$\\ 
\hline
conv\_8e6 & $8 \times 10^6$ & $\mathbf{1.0}$ &  $\mathbf{no}$\\ 
\hline
conv\_1e7 & $10^7$ & $\mathbf{1.0}$ &  $\mathbf{no}$\\ 
\hline
conv\_5e7 & $5 \times 10^7$ & $\mathbf{1.0}$ &  $\mathbf{no}$\\ 
\hline
conv\_1e9 & $10^9$ & $\mathbf{1.0}$ &  $\mathbf{no}$\\ 
\hline
\end{tabular}
\caption{Characteristics of our RT simulation runs. From left to right: name; number of photon packets emitted from each stellar particle in the RT simulation; local (i.e. subgrid) escape fraction; inclusion of dust as an additional absorber of ionizing radiation. Bold entries refer to the fidcuial configuration.
}
\label{tab:configs} 
\end{table}

When the RT simulation is run, each source emits $N_p$ photon packets, each containing
\begin{equation}
N_j = \frac{\dot{N}_j t_s}{N_p}
\end{equation} 
individual photons distributed over a given number of frequency bins, with the distribution being determined by the SEDs of the sources. Here $t_s$ is the total time of the RT simulation. As a photon packet crosses a given grid cell, the optical depth is calculated as
\begin{equation}
\tau_{i,\nu} = n_i \sigma_{i,\nu} l, \label{optical depth}
\end{equation}
where $\nu$ is the frequency bin, $l$ is the length of the path through that grid cell, $n_i$ is the local density of species $i$ and $\sigma_{i,\nu}$ its cross section in that frequency bin. The number of absorbed photons in a grid cell and frequency bin is then given by:
\begin{equation}
\Delta N_\nu = N_\nu(1-\mathrm{e}^{-\tau_\nu}) \hspace{0.5cm}\mathrm{with}\hspace{0.5cm} \tau_\nu = \sum_i \tau_{i,\nu},
\end{equation}  
where $N_\nu$ is the number of photons in the frequency bin $\nu$. As a result, the physical state of the gas is updated in terms of its temperature and ionization state (we refer the reader to the original papers for more details). This process is repeated until the photon packet leaves the simulated volume or, in the case of periodic boundary conditions, until the number of photons left in the packet drops below some threshold.

When taking into account dust absorption, the optical depth is further modified by the contribution of dust to the absorption cross section in the same fashion as for all other gas species. The crucial difference here however is that there is no tracking of the ionization state of the dust, i.e. the photons are simply absorbed by the dust without changing its physical properties.

\subsection{Running CRASH on TNG50 galaxies}

We first map the gas distribution of a given TNG50 halo onto a uniform grid. To do so, we derive an adaptive smoothing length for each Voronoi gas cell in the TNG50 simulation, equal to the radius of the sphere enclosing its 64 nearest neighbors. We then distribute the physical properties of each gas cell onto a regular Cartesian grid, where the deposition weighting uses the usual SPH cubic-spline kernel. The physical size of the grid is chosen such that each side has a length of twice the virial radius of the halo ($R_{\mathrm{vir}}$). The required size of the grid cells is determined by the minimum smoothing length of the particles to ensure that the resolution of our CRASH runs matches the resolution of TNG50. As the values for $x_{\mathrm{HeII}}$ and $x_{\mathrm{HeIII}}$ are not directly provided by the TNG snapshots, we calculate them for each cell based on collisional ionization equilibrium for a given gas temperature and density.

Note that we set star-forming gas cells from TNG to be initially neutral and to have a temperature of $1000$K. This choice is consistent with the two-phase ISM model of the hydrodynamical simulations, where the dominant fraction ($> 0.9$) of gas mass is contained in the cold phase \citep{springel03}.

For the sources of radiation in the RT simulation we use the stellar particles from the TNG50 simulation, which are representations of stellar populations. To obtain the spectra and emissivites of these particles we use the BPASS library \citep{bpass1, bpass2}, based on the age, metallicity and mass of each stellar particle. There is no significant contribution of ionizing radiation emitted from AGN for galaxies within our examined mass range, and we omit any AGN contribution in our RT calculations.

With the gas of the halo distributed on a regular grid, together with the radiation sources and their spectra, we then run the CRASH simulation. Our fiducial configuration is shown in Table \ref{tab:configs}, together with several variations on both physical and numerical parameters which we also explore in this work. Each CRASH run extends for $5$\,Myr, a time span which guarantees that by the end of the simulation ionization equilibrium is achieved, although typically this is reached even earlier, within $\lesssim 1$ Myr. We use $N_p=10^8$ for our studies to guarantee convergence (see Appendix \ref{app:conv} for a convergence analysis). Each packet is subdivided into 64 frequency bins sampling photon frequencies in the range $(13.6-161.9)$\,eV, where the upper energy limit is chosen to capture helium ionization and explore the impact of higher energy photons emitted mainly by binary stars.

The full sample of halos summarized in Figure \ref{fig:sample} is processed in our fiducial model in which we set the `local' (i.e. unresolved) escape fraction $f_\mathrm{esc,loc}$ to unity, and do not include the absorption effects of dust. This choice for the local escape fraction was adopted as the simplest choice for a largely unconstrained parameter, and we discuss this value further in Section \ref{sec:fescloc}.

\subsection{Photon Escape Fraction}

\begin{figure*}
    \centering
    \includegraphics[width=\textwidth]{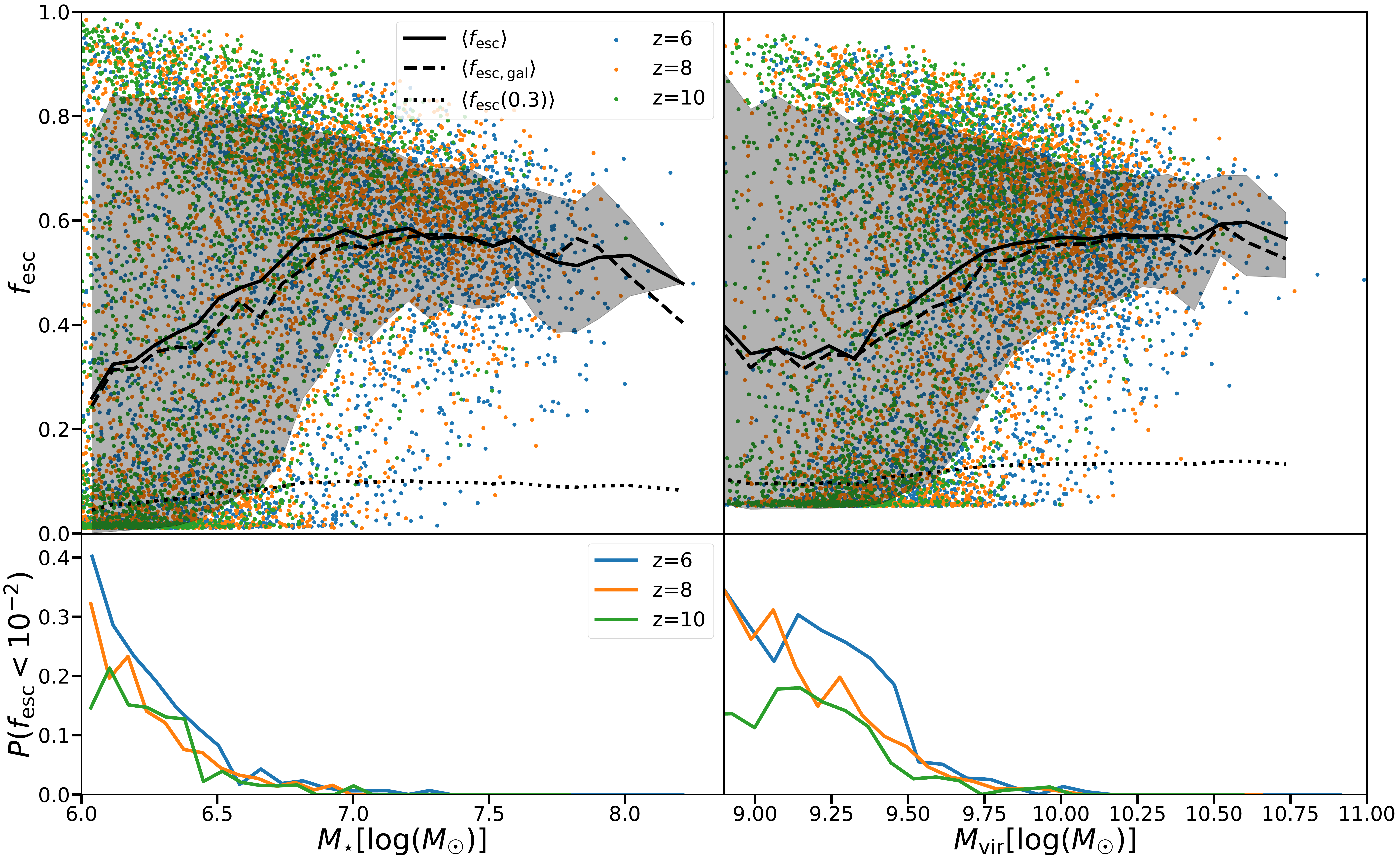}
    \caption{{\it Top panels:} The escape fraction of ionizing radiation as a function of galaxy stellar mass (left panels) and total halo mass (right panels) at redshifts $z=6$ (blue), $z=8$ (orange) and $z=10$ (green), as predicted by our TNG50 post processing with CRASH. The solid and dashed lines refer to the average halo and galactic escape fractions, which increase with mass from $\sim\,0.3$ at $M_\star \sim 10^6$\,M$_\odot$ to a maximum of $\sim\,0.6$ at $M_\star \sim 10^8$\,M$_\odot$, after which there is a hint of a turnover and a decrease of escape fractions for even more massive galaxies. The dotted line shows our results rescaled to $f_{\rm esc,loc}=0.3$, instead of our fiducial choice of unity, resulting in significantly lower escape fractions (see text). Shaded areas represent the interval containing $68\%$ of all halos. 
    {\it Bottom panels}: The fraction of halos with an negligible escape fraction (below $10^{-2}$), as a function of mass. The abundance of such non-contributing galaxies rises rapidly towards the lowest masses considered in our study, due to the significant stochasticity in the stellar populations of low-mass galaxies.}
    \label{fig:med_fesc}
\end{figure*}

To calculate the halo escape fraction from the TNG50 simulation we proceed in the following way. First, we cast photon packets from each source in random directions and trace them until they reach the virial radius, $R_\mathrm{vir}$, of the halo. The number of escaped photons in the frequency bin $\nu$ of one photon packet is given by 
\begin{equation}
N_{\nu,\mathrm{esc}} = N_{\nu, \mathrm{em}} \times \mathrm{e}^{-\tau_{\nu, \mathrm{tot}}},
\end{equation}
where $N_{\nu, \mathrm{em}}$ is the number of emitted photons in a packet with frequency bin $\nu$, and $\tau_{\nu, \mathrm{tot}}$ is the optical depth along the path of the ray, sampling each intersected gas cell using eqn.~\eqref{optical depth}. For each stellar particle we initially cast $1000$ packets. The total escape fraction from the halo is then be calculated as
\begin{equation}
\mathrm{ \mathnormal{f}_{esc} = 
  \frac{\sum_{source}\sum_{packet}\sum_\nu \mathnormal{N}_{\nu,esc}(source,packet)}
       {\sum_{source}\sum_{packet}\sum_\nu \mathnormal{N}_{\nu,em} (source,packet)}
}.
\end{equation}
We repeat the calculation of $f_\mathrm{esc}$ until we reach convergence to better than an absolute tolerance of $10^{-2}$.


\section{Results} \label{sec:fesc}

\subsection{Halo Escape Fraction}

We begin with our main result. Figure \ref{fig:med_fesc} shows the escape fraction of ionizing photons as a function stellar mass (left) and total halo mass (right). This is the halo-scale escape fraction, and overall we observe a large scatter at all redshifts and masses, with the variation increasing for lower mass galaxies. This scatter is largely driven by the stochasticity of small stellar populations within the lower mass halos. In such halos, in fact, $f_{\rm esc}$ can be dominated by the escape fraction of a handful of star particles, which in turn strongly depends on the age of the population. Indeed, young stellar populations produce significant amounts of ionizing radiation, and as the cores of smaller halos are less dense, a young stellar population is usually able to ionize its surrounding.

In larger halos, instead, the escape fraction is more strongly dominated by the global properties of the halos, as random fluctuations in the escape fraction of individual stellar particles are averaged out by having a sufficiently large sample. As a result, the distribution of the escape fraction becomes bimodal at small masses, where a large number of halos have high escape fractions, while a significant fraction also have $f_\mathrm{esc}<1\%$ and therefore do not contribute significantly to the ionizing photon budget escaping into the IGM. As these halos are difficult to distinguish in the top panel, we plot the fraction of halos with $f_{\rm esc}<1\%$ in the lower panel. Indeed, the fraction of halos with $f_{\rm esc}<1\%$ is $\approx 30\%$ for small mass halos, while it becomes negligible for halos with stellar masses $\gtrsim 10^{6.5}$\msun.

\begin{figure*}
   \centering
   \includegraphics[width=\textwidth]{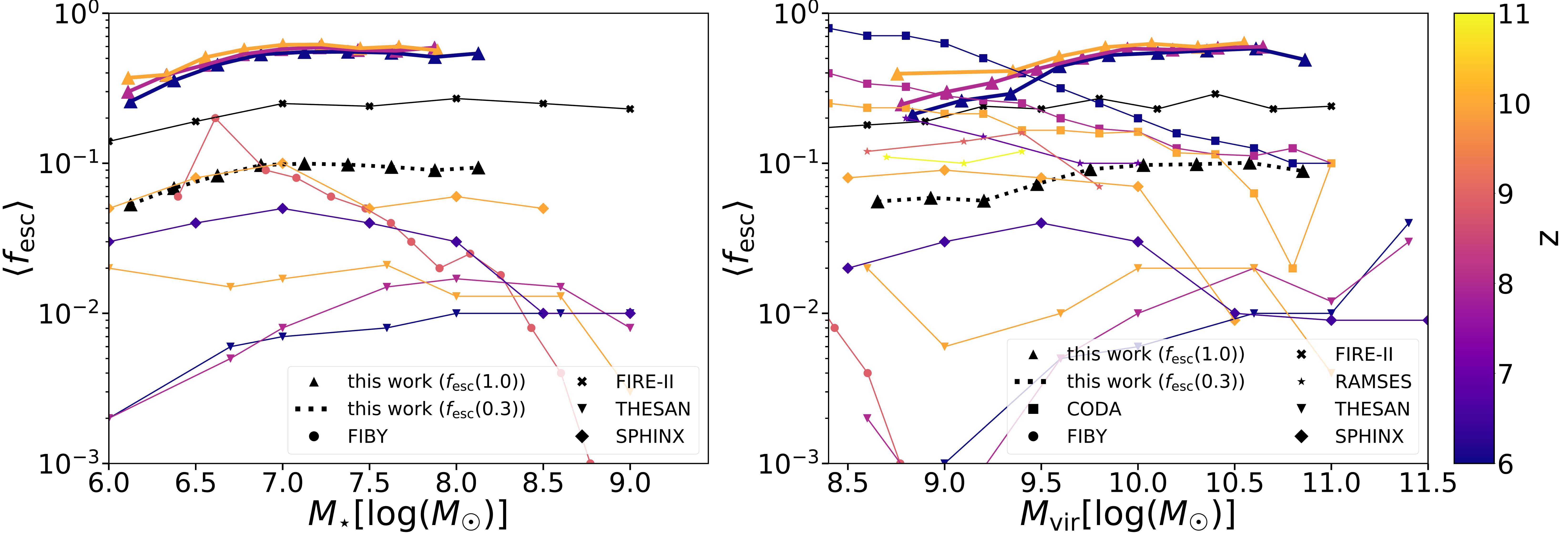}
   \caption{Comparison of escape fractions between different simulations, as a function of galaxy stellar mass (left) and halo mass (right). We contrast our TNG50+CRASH RT simulations (this work, $f_\mathrm{esc}=1.0$, thick lines with large triangles), as well as our result rescaled to $f_\mathrm{esc,loc}=0.3$ (dotted black line with large black triangles), with results from FIRE-II with halos across $z=5-12$ \citep{ma20}, the First Billion Years project \citep{paardekooper15}, the Cosmic Dawn II simulation \citep[CoDa-II;][]{lewis20}, the THESAN simulation \citep{yeh22}, the RAMSES simulation by \citet{kimm14}, as well as the SPHINX simulation of \citet{rosdahl22}. Note that different calculations have adopted different sub-grid escape fractions. The THESAN simulations assume a value of $f_\mathrm{esc,loc}=0.37$, the CoDa-II uses $f_\mathrm{esc,loc}=0.42$, while the others all assume $f_\mathrm{esc,loc}=1$. There is significant diversity in the results, and predictions for the escape fraction of ionizing radiation vary by at least an order of magnitude, at any given mass, between different simulations.}
    \label{fig:fesc_comparison}
\end{figure*}

The average escape fraction, $\langle f_{\rm esc} \rangle$, calculated by combining all redshifts, initially increases with increasing mass. It then flattens at $M_{\star} = 10^7$\msun and for $M_{\star} \gtrsim 10^8$\msun there is an indication that the escape fraction begins again to decrease, although here our sample of RT-processed halos is small. The same trend is observed at each individual redshift, where the average escape fraction slightly decreases with decreasing redshift (see Figure \ref{fig:fesc_comparison}).

In contrast to the total escape fraction (solid lines), the galactic escape fraction $\langle f_{\rm esc,gal} \rangle$ (dashed lines) is defined as the relative amount of photons that escapes a sphere of radius $R_\mathrm{gal}=0.2R_\mathrm{vir}$, representing an approximate boundary between the galaxy and its circumgalactic medium \citep{nelson13}. Although the evolution of $\langle f_{\rm esc,gal} \rangle$ is similar to that of $\langle f_{\rm esc} \rangle$, its value is on average slightly lower. This occurs because stellar particles located outside $R_\mathrm{gal}$ are embedded in a less dense environment, which in turn facilitates the escape of ionizing photons.

In addition to this fiducial result, we have also included the escape fraction we would have obtained with a sub-grid local escape fraction of $f_\mathrm{esc,loc}=0.3$ (dotted lines), as this value leads to results which are in better agreement with observational constraints on global emissivities. The range between these lines represents an element of systematic uncertainty on the emergent escape fractions from the cosmological simulation, and this range is significant. We further discuss the important role of $f_\mathrm{esc,loc}$ in Section \ref{sec:fescloc}.

In Figure \ref{fig:fesc_comparison} we compare our results against other simulation-based inferences. We show escape fraction as a function of stellar mass (left) and halo mass (right), across the redshift range of reionization (as indicated by the colors). The previous studies with published results we compare against are: the FIRE-II simulation \citep{ma20}, the First Billion Year project \citep[FiBY;][]{paardekooper15}, the CoDa-II simulation \citep{lewis20}, the RAMSES simulations by \cite{kimm14}, the THESAN simulation \citep{yeh22}, and the SPHINX simulation \citep{rosdahl22}.\footnote{Note that \citet{rosdahl22} used a luminosity weighting for the calculation of the average escape fraction.}

Broadly speaking, predictions for the escape fraction of ionizing radiation during reionization span an enormous range. Between different simulations, $f_{\rm esc}$ easily varies by an order of magnitude, at fixed mass. There is clearly no theoretical consensus on the average value of this parameter, nor its detailed relationship with galaxy mass, halo mass, or redshift.

The qualitative trends of our $f_{\rm esc}$ (discussed above) agree well with those from the FIRE-II simulations (black crosses), which adopt the same approach to the problem, i.e. the radiative transfer is based on Monte Carlo post-processing. On the other hand, the normalization of our fiducial results is higher possibly due to the different resolution of the simulations. 

A more realistic value of $f_{\rm esc,loc}$ results in a reduction of $f_{\rm esc}$ (see Figure \ref{fig:unresolved}), as can be seen from the dotted line in the figure, which shows the average escape fraction rescaled to the value it would have with a sub-grid local escape fraction of $f_\mathrm{esc,loc}=0.3$.

The CoDa-II RHD simulations \citep{lewis20} show a stronger dependence on the halo mass (squares), where the trend at the high-mass end is sensitive to redshift. The direction of the correlation with mass is qualitatively inconsistent with our results in the overlapping mass range, although the actual magnitude of $f_{\rm esc}$ is similar. The FiBY results \citep{paardekooper15}, also based on a post-processed RT, show a similarly strong decreasing trend of escape fraction with stellar mass, inconsistent with our findings. The SPHINX results \citep[diamonds;][]{rosdahl22} exhibit qualitatively similar trends as our TNG50 results, albeit at a normalization slightly below our $f_\mathrm{esc,loc}=0.3$ case. The \cite{kimm14} simulations (stars) have qualitatively similar results as this work, bracketed by the two cases shown, albeit over a limited mass range.

The most compelling comparison is with the escape fractions predicted by THESAN \citep[downward triangles;][]{yeh22}, which are by and large much lower than ours, with typical values of order $\sim\,1$\%, with mass trends which are either flat, slightly increasing, or slightly decreasing, depending on redshift. That work finds a clear signature of decreasing $f_{\rm esc}$ at the high-mass end, for $M_\star \gtrsim 10^{8.5}$\,M$_\odot$, where our current sample runs out of statistics.

The level of disagreement with THESAN is notable because it is based on the same IllustrisTNG galaxy formation model, essentially unchanged from its fiducial incarnation in TNG50, as we analyze here, except for the notable addition of on-the-fly radiative transfer \citep{kannan19}. In THESAN, a non-trivial complexity therefore arises in how radiation interacts with star-forming gas \citep{smith22}. We remind the reader that in the TNG model this gas has its thermal state set by a sub-grid, two-phase model for pressurization of the ISM by unresolved stellar feedback processes \citep{springel03}. Because the effective temperature is high, dense ISM gas is effectively ionized, and the escape fraction from the (resolved) ISM in THESAN is therefore unity. Different choices in the coupling of the RT with the sub-grid ISM model could change these results, and our comparison. The on-the-fly RT treatment in THESAN is clearly a key strength, and captures the dynamic impact and back-reaction on the gas, which we miss in TNG50. However, our post-processing approach also enables a more accurate (and expensive) Monte Carlo RT solution. Finally, the hydrodynamical resolution of TNG50 is significantly higher than THESAN -- eight times better in baryonic mass resolution, and this could also play a role in resolving small-scale structure in the ISM. A full understanding of the predictions for escape fractions in the TNG galaxy formation model will benefit from a future, apples-to-apples comparison between our current results and those from THESAN.

With respect to redshift trends, we find that the average $f_{\rm esc}$ decreases with decreasing redshift, although the trend is weak, from a few percent to perhaps ten percent from $z=10$ to $z=6$. We note that this correlation with redshift is expected for models which produce enough photons to sustain the reionization process \citep{fujita2003,razoumov10}. The CoDa-II and \cite{kimm14} simulations show the opposite trend, while the FiBY simulation finds no strong dependence, and THESAN suggests a non-monotonic behavior.

Although it is difficult to ascertain the reasons for such discrepancies, we note that the escape fraction is strongly dependent on the small scale density distribution of gas in the ISM, and hence on simulation resolution, which varies (substantially) between all the above studies. While we investigate the effect of the unresolved ISM on $f_{\rm esc}$ by varying the local escape fraction (see Figure \ref{fig:unresolved}), we assume that its value is universal and constant: independent of halo mass and redshift, for example. Such dependencies could change the behaviour of $f_{\rm esc}$, in particular for larger halos, as we discuss in Section \ref{sec:fescloc}.

\begin{figure}
    \centering
    \includegraphics[width=0.225\textwidth]{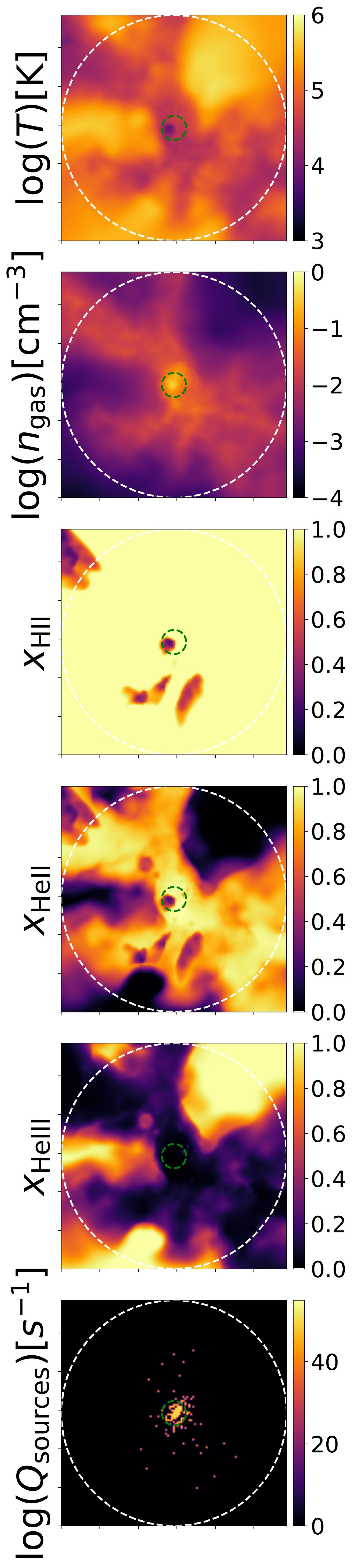}
    \includegraphics[width=0.225\textwidth]{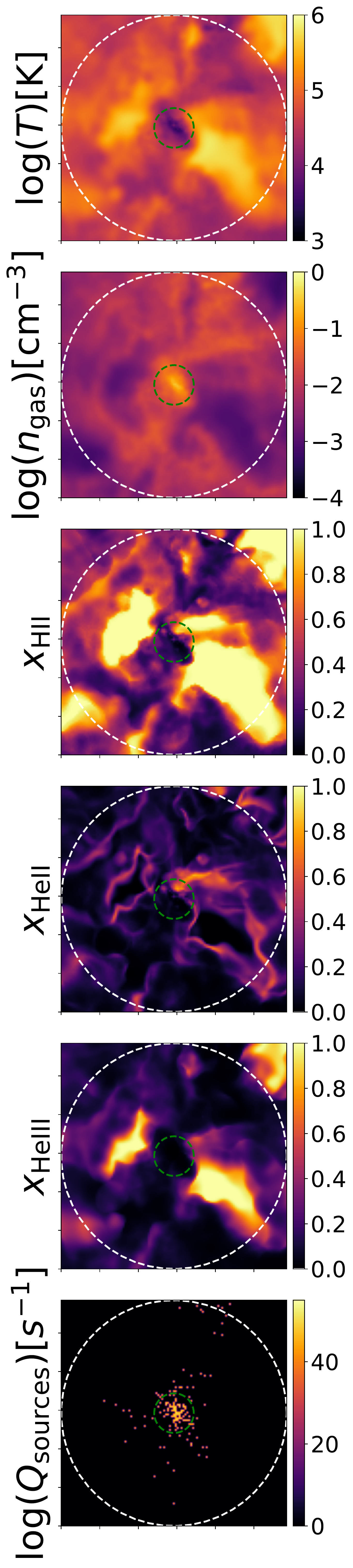}
    \caption{Two similar mass halos with different escape fractions. From top to bottom, the panels show 2D projections of temperature, gas number density, HII, HeII and HeIII fractions and stellar particle emissivity after the RT simulation run. The left (right) panels correspond to a $z=8$ halo with mass $M_\mathrm{vir}=5.5\times10^{9}$\msun ($8.2\times10^{9}$\msun), and $f_{\rm esc} = 76.5$\% (3.6\%). The white and green circles indicate the virial radius of the halo, and the radius encompassing 50\% of the total stellar particle emissivity, respectively. The temperature and the ionization states are density weighted in projection. 
    }
    \label{fig:halo_image}
\end{figure}

In Figure \ref{fig:halo_image} we go deeper into the gas physics relevant for the escape fraction calculation, visualizing two representative $z=8$ halos with similar mass, but with very different $f_{\rm esc}$. Indeed, the halos in the left and right panels have a mass of $M_\mathrm{vir}=5.5\times10^{9}$\msun and $8.2\times10^{9}$\msun, respectively, while their escape fractions are $76.5\%$ and $3.6\%$, respectively. As a result, these two halos fall at the upper and lower end of the scatter in $f_{\rm esc}$ for their mass.

We create two-dimensional projections of gas temperature, $T$, the gas number density, $n_{\rm gas}$, the fractions of HII, $\mathrm{x}_\mathrm{HII}$, HeII, $x_\mathrm{HeII}$, and HeIII, $x_\mathrm{HeIII}$, and the stellar particle emissivity, $Q_\mathrm{sources}$. In both halos we see that the central gas cloud, at high density, remains cold and mostly neutral. At larger distance, the right-side halo exhibits neutral gas in appreciable quantities along overdense filaments and extended structures. In contrast, the left-side halo is comprised of fully ionized hydrogen, with the exception of a few localized clouds of gas which manage to remain mostly neutral. 

As expected, the vast majority of ionizing sources are located in the center of the halo, where the central galaxy is found. The halo on the left has an ionization rate of $6 \times 10^{52}$~s$^{-1}$, while the one on the right of $2 \times 10^{51}$~s$^{-1}$. This difference in ionizing photon production rate, i.e. the scatter of stellar population properties at a given halo mass, is the key driver of the scatter in escape fraction. 

In addition, we see that in the left-hand side halo the overall source distribution is somewhat displaced (to the right) with respect to the central overdense region, in the right-hand side halo they are mostly aligned with the halo center. In the former case, the radiation can ionize the surrounding gas more easily than in the latter, an effect which can be seen in the $x_\mathrm{HII}$ maps.

In the left-hand side halo the sources are also able to ionize the surrounding HeI. However, in both halos $x_\mathrm{HeIII}$ is negligible in the central region and the only HeIII present is due to ionization from shocks. We therefore expect that the higher energy tail of the radiation emitted by the sources will be absorbed more in comparison to the less energetic photons (see discussion in Section \ref{sec:spectra}). 

\begin{figure*}
    \centering
    \includegraphics[width=0.49\textwidth]{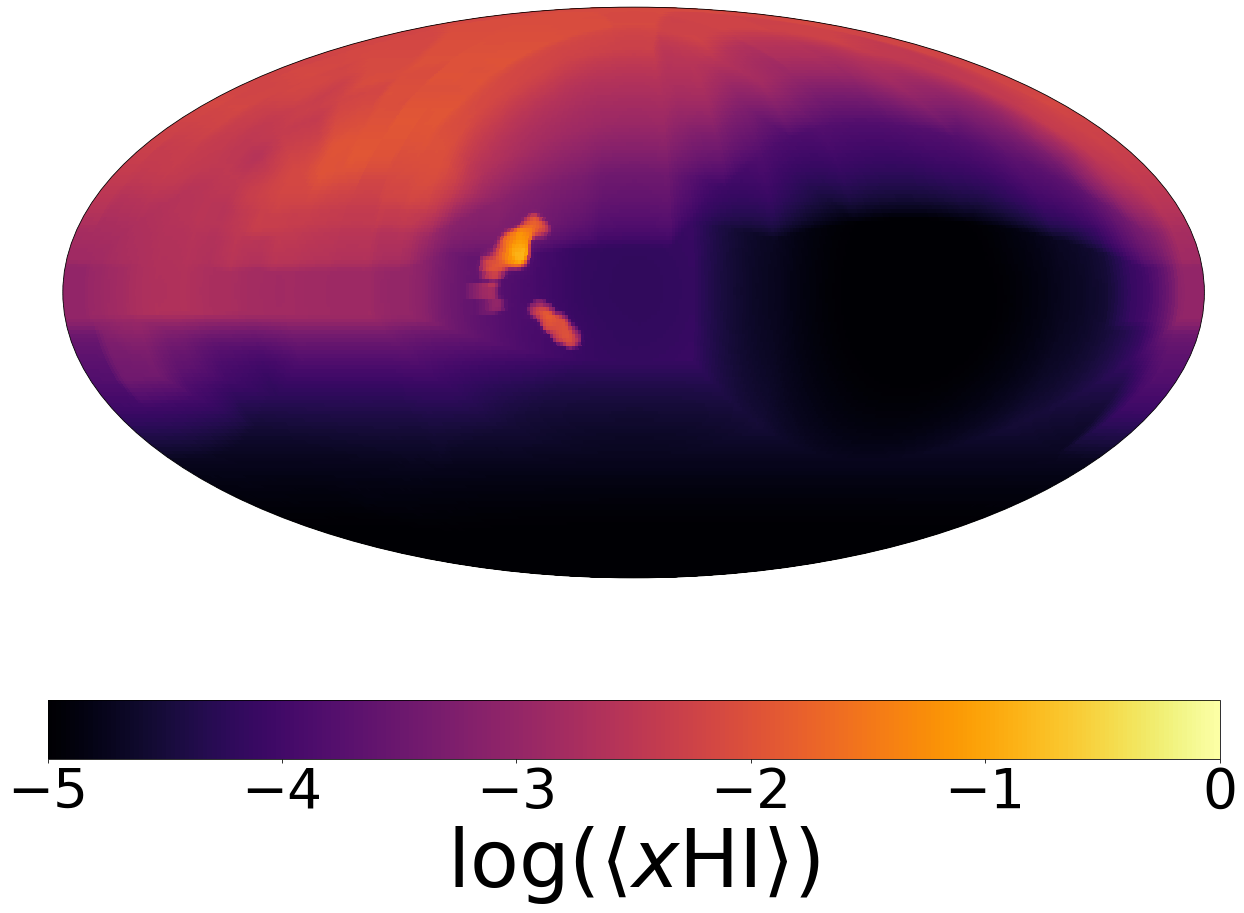}
    \includegraphics[width=0.49\textwidth]{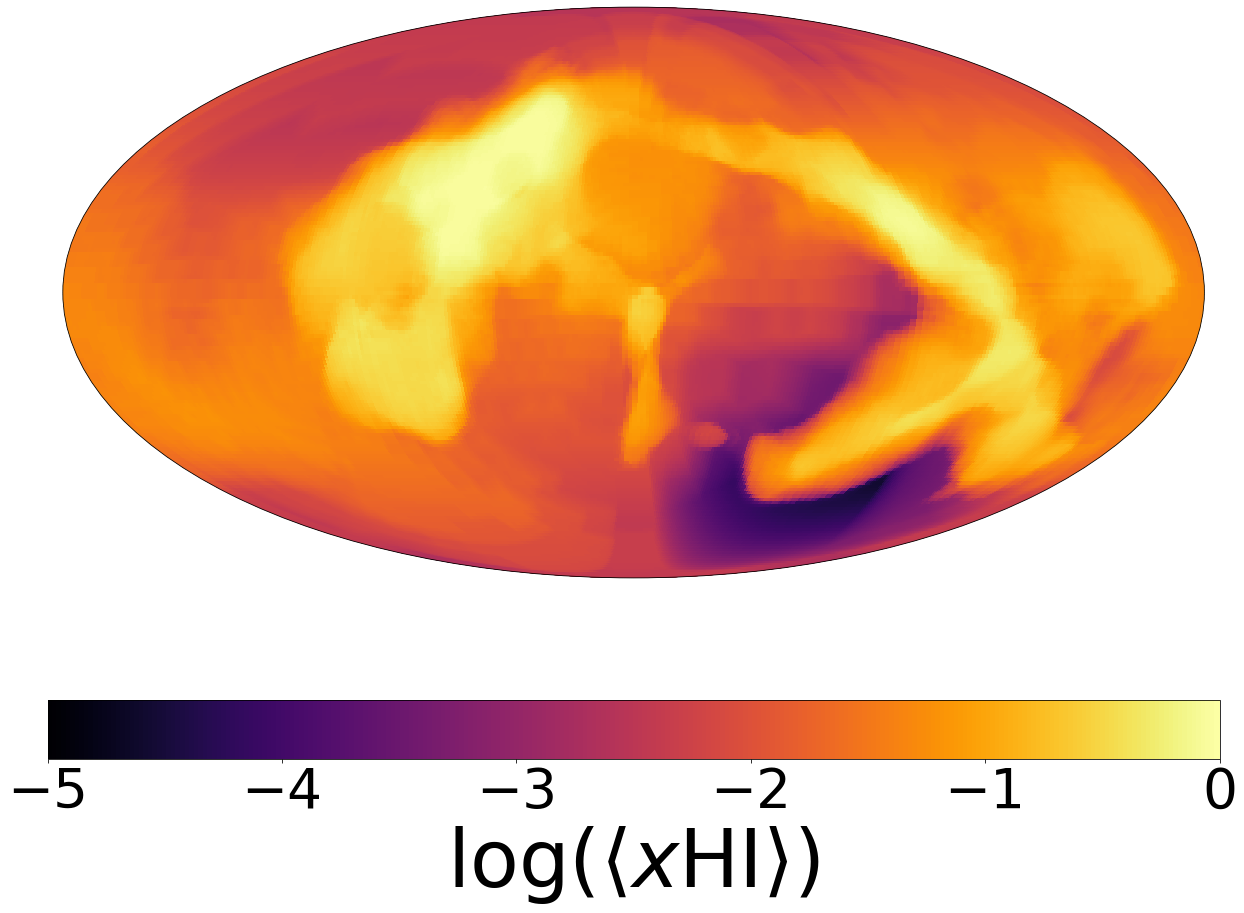} \\
    \includegraphics[width=0.49\textwidth]{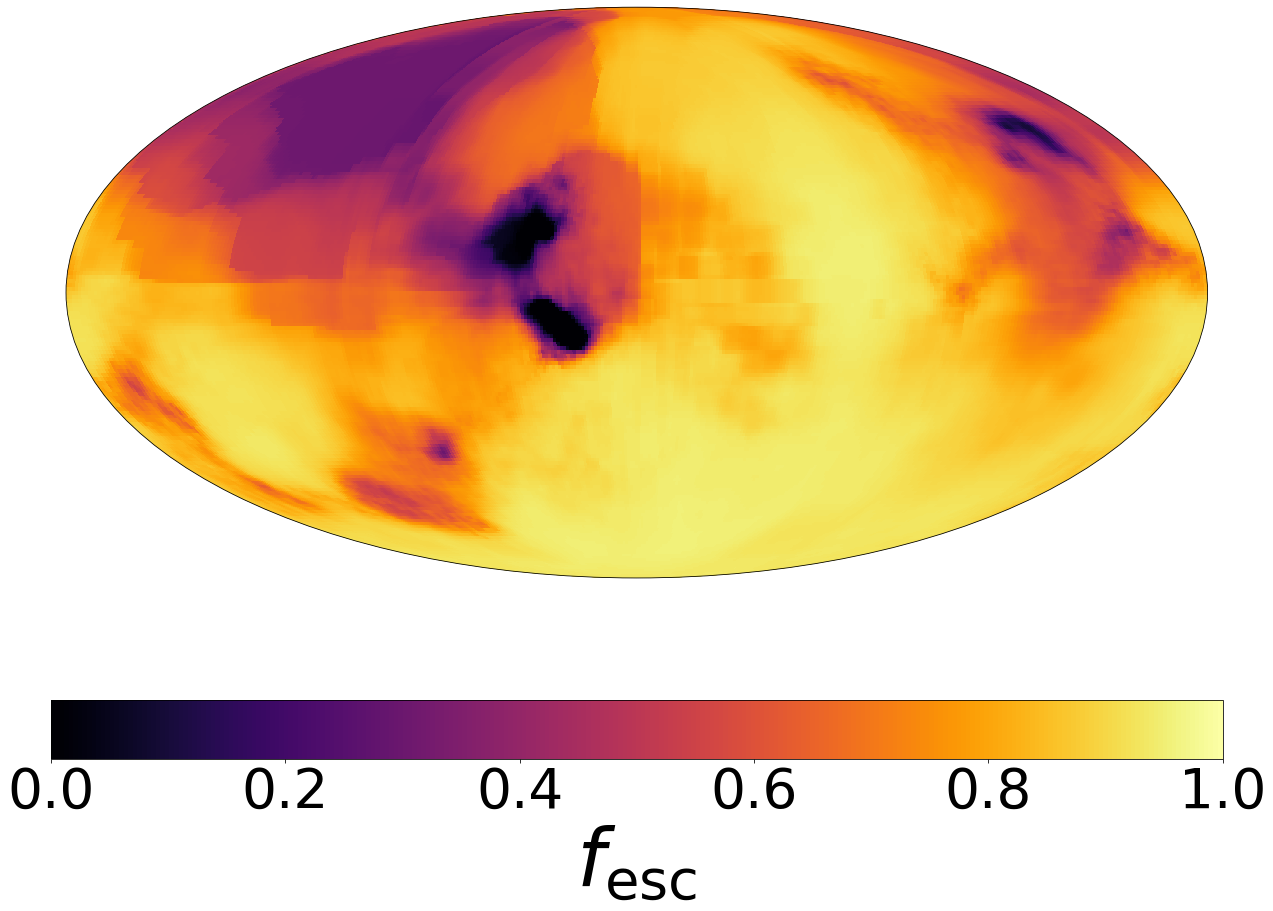}
    \includegraphics[width=0.49\textwidth]{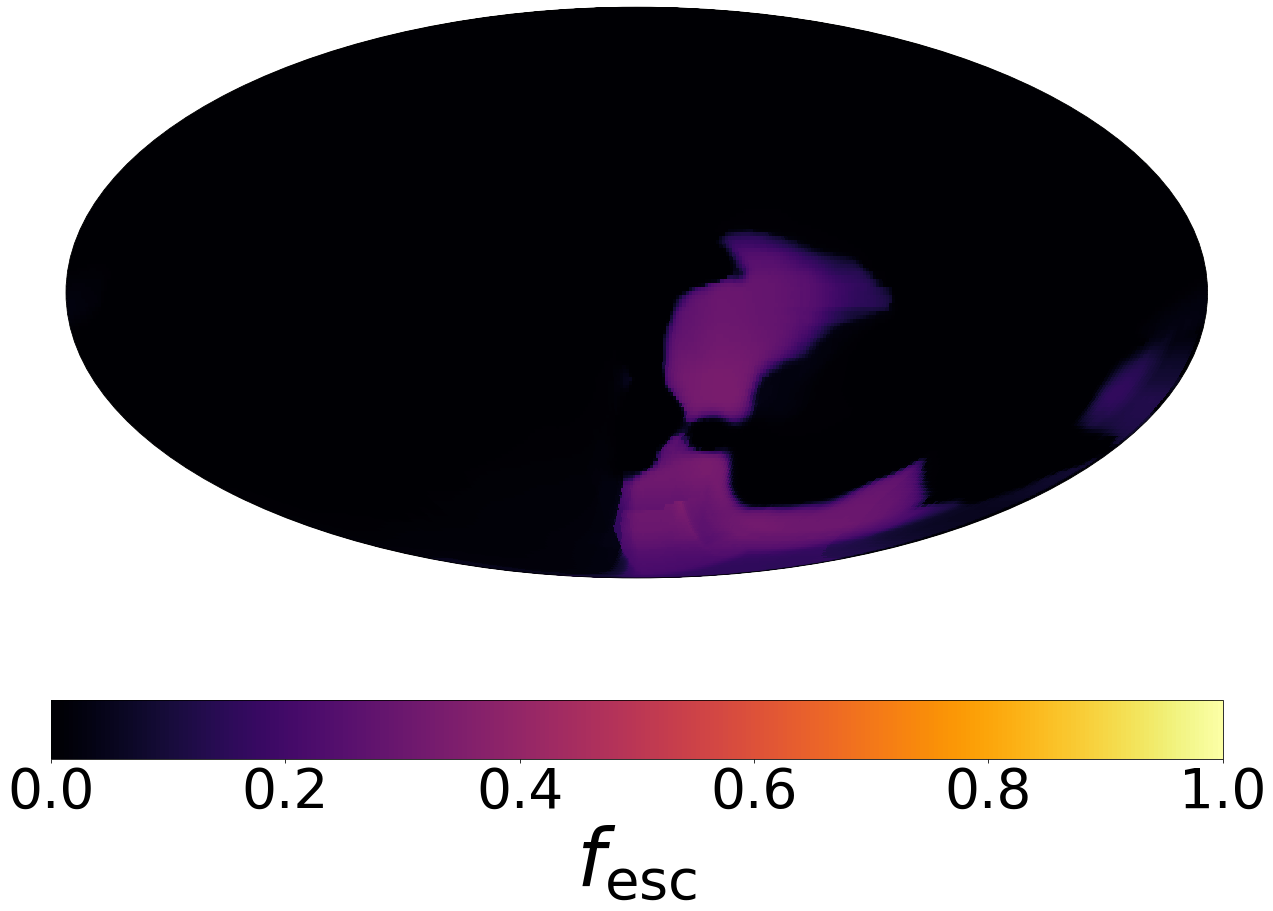}
    \caption{Mollweide all-sky projections of the same two halos presented in Figure \ref{fig:halo_image}. \textit{Top:} Average neutral hydrogen fraction along radial lines of sight from the halo center to a sphere at the virial radius. \textit{Bottom}: Emissivity weighted escape fraction of all sources at the virial radius of the halo, i.e. the fraction of photons emitted from all sources in a given direction which reach the virial radius of the halo.}
    \label{fig:spherical_proj}
\end{figure*}

In Figure \ref{fig:spherical_proj} we investigate in more detail the spatial dependence of the escape fraction in the same two example halos. Here we show the average neutral fraction of hydrogen (top), together with the corresponding average emissivity weighted escape fraction (bottom), along radial lines of sight from the halo center to the virial radius. As expected, there is an anti-correlation between the escape and average $x_{\rm HI}$, with photons preferentially escaping through channels of circumgalactic gas which are highly ionized.

In the halo on the left ($f_{\rm esc}=76.5\%$) photons are able to escape through most of the virial surface of the halo, the only exception being small regions of solid angle with residual dense and neutral gas at the center and in the top-left corner, also visible in Figure \ref{fig:halo_image}. In the halo on the right ($f_{\rm esc}=3.5\%$) only a relatively small area in the bottom-right part of the image has a significant escape fraction, corresponding to channels of the most ionized gas, visible in the top panel as well as in Figure \ref{fig:halo_image}. The escape fraction through this channel, though, is fairly low  ($\approx 30\%$) since a significant fraction of the stellar particles are located outside of it, such that ionizing photons emitted from these particles are largely absorbed.

The escape of ionizing radiation depends strongly on direction.

\subsection{Dependence on galaxy and halo properties}

In order to explore the variation of $f_\mathrm{esc}$, we study its dependence on a number of properties of both the central galaxy and its parent dark matter halo. We note that the phrases `within the galaxy/galactic' or `halo' refer to the collection of gas cells or particles residing within $R_\mathrm{gal} = 0.2 R_{\rm vir}$ or $R_\mathrm{vir}$, respectively. All properties are calculated directly from the CRASH grids, to guarantee they are consistent with the RT inputs. Finally, as these dependences of $f_\mathrm{esc}$ on different properties have little redshift dependence, we combine all halos across redshifts in the following analysis.

In Figure \ref{fig:color1} we show eight panels, each visualizing the plane of escape fraction versus stellar mass. In each case, we color bins by the median value of a third physical property, as indicated by the colorbars and labels. In addition, we normalize these physical values by their median at each given stellar mass bin, making the color scales relative (i.e. conditional) to the typical property of galaxies or halos at that mass. To represent the actual density of objects in this plane, we use the solid and dashed contours to demarcate the $1\sigma$ and $2\sigma$ regions of galaxy occupation.

\subsubsection*{Gas surface density}

First we examine the effect of galactic gas surface density, defined as

\begin{equation}
    \Sigma_{\mathrm{gas, gal}} = M_\mathrm{gas, gal} / \left( \pi R_\mathrm{gal}^2 \right) \label{eq:surface_density},
\end{equation}
where $M_{\mathrm{gas, gal}}$ is the gas mass within the galactic radius $R_\mathrm{gal}$ (top left panel). We observe two effects in the escape fraction due to variation in $\Sigma_{\mathrm{gas, gal}}$. First, higher gas surface density leads to a higher SFR and hence more stars which can ionize the surrounding medium and create channels for ionizing radiation to escape. On the other hand, higher gas surface density also implies that the medium is more opaque, making it more difficult to ionize. We find that the second effect dominates, as the escape fraction negatively correlates with the surface density of the gas. For example, at $M_\star = 10^7$\,M$_\odot$, the escape fraction decreases from $\sim\,0.8$ to $\sim\,0.6$ as the gas surface density increases by a factor of two. At the same time, outliers with low $f_\mathrm{esc} \lesssim 0.4$ at this mass scale exhibit similarly low gas surface densities as the highest escape fraction galaxies. Such a non-monotonic trend is visible at intermediate galaxy masses.

\subsubsection*{Stellar surface density}

\begin{figure*}
    \centering
    \includegraphics[width=0.43\textwidth]{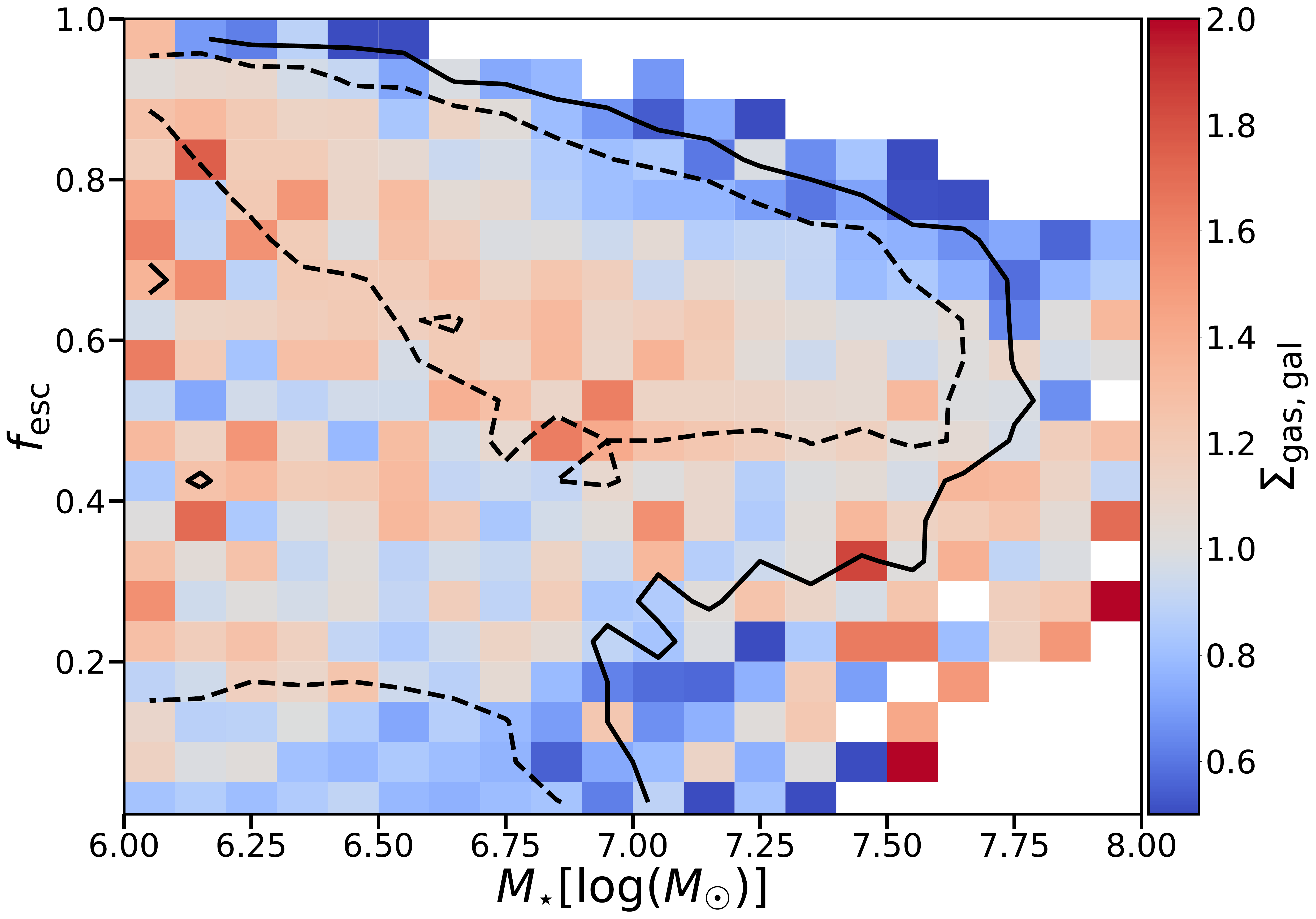}
    \includegraphics[width=0.43\textwidth]{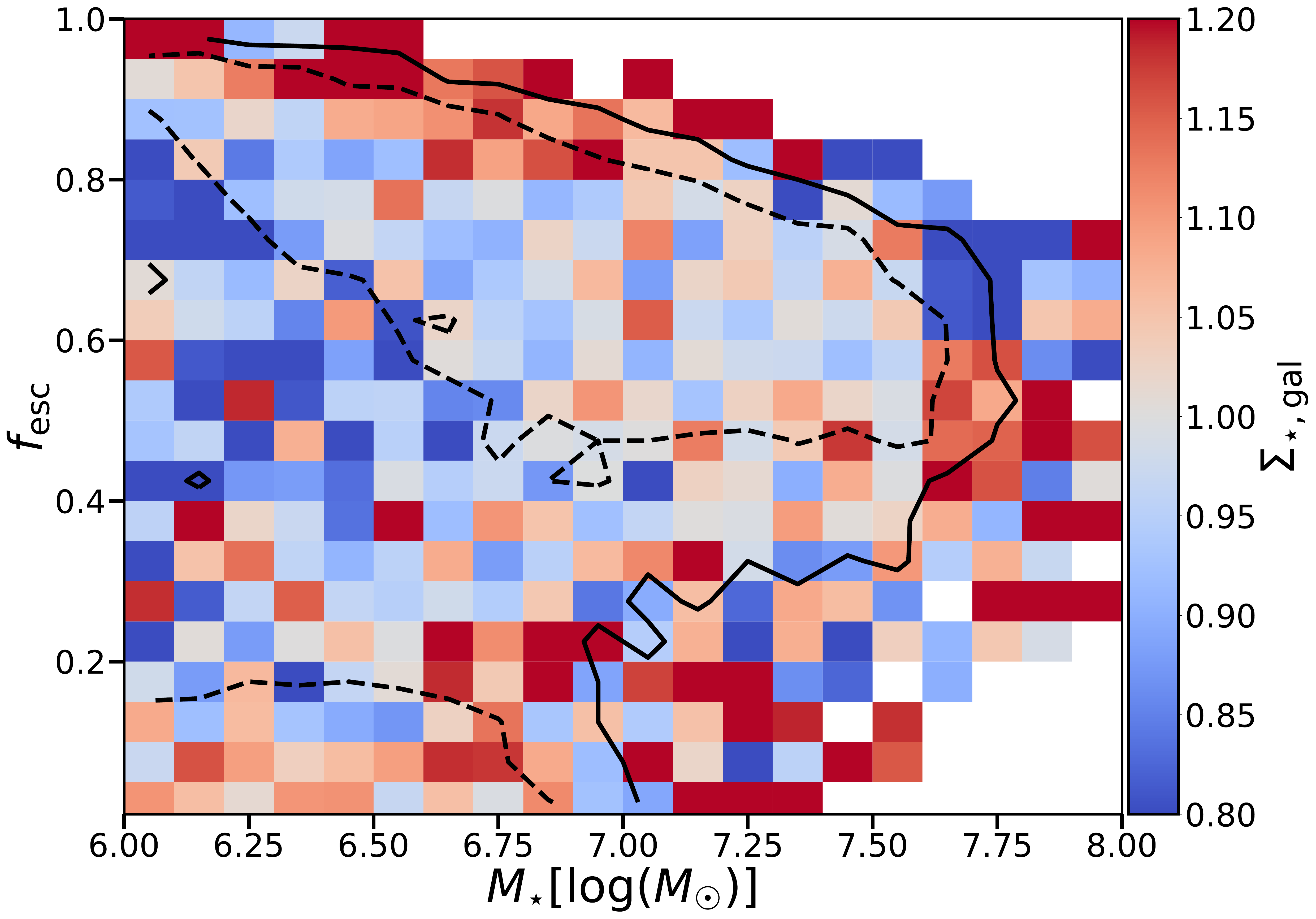}
    \includegraphics[width=0.43\textwidth]{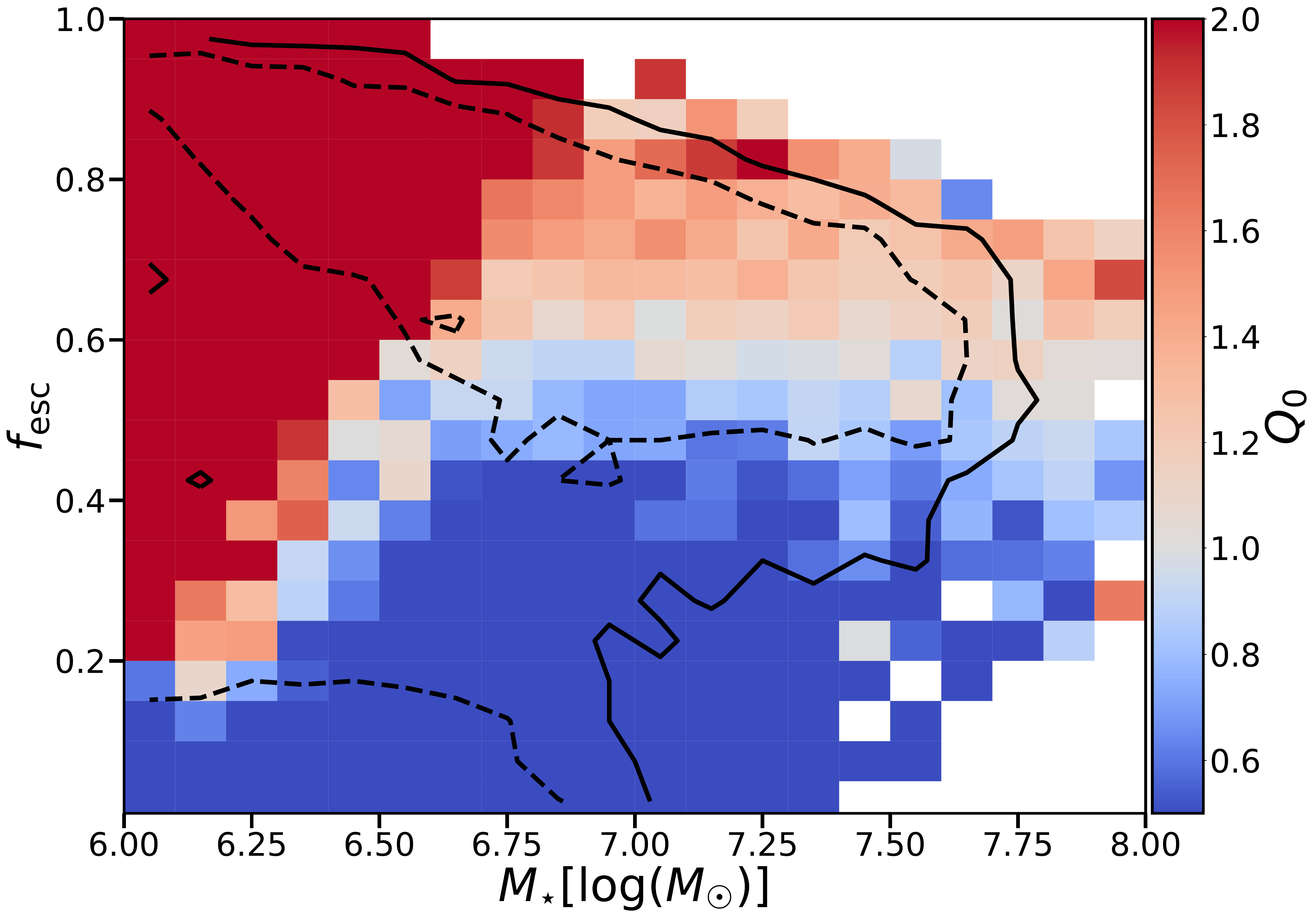}
    \includegraphics[width=0.43\textwidth]{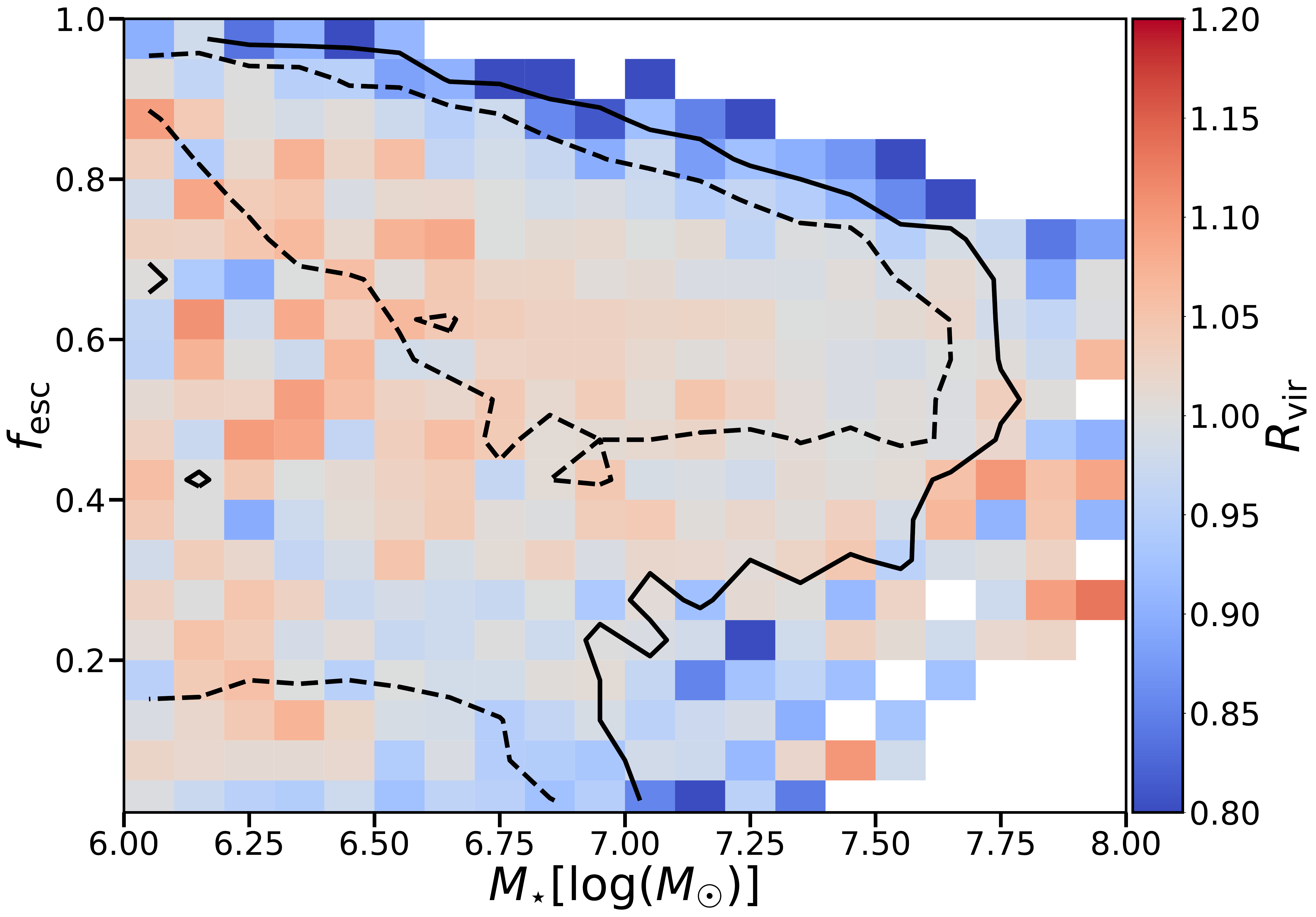}
    \includegraphics[width=0.43\textwidth]{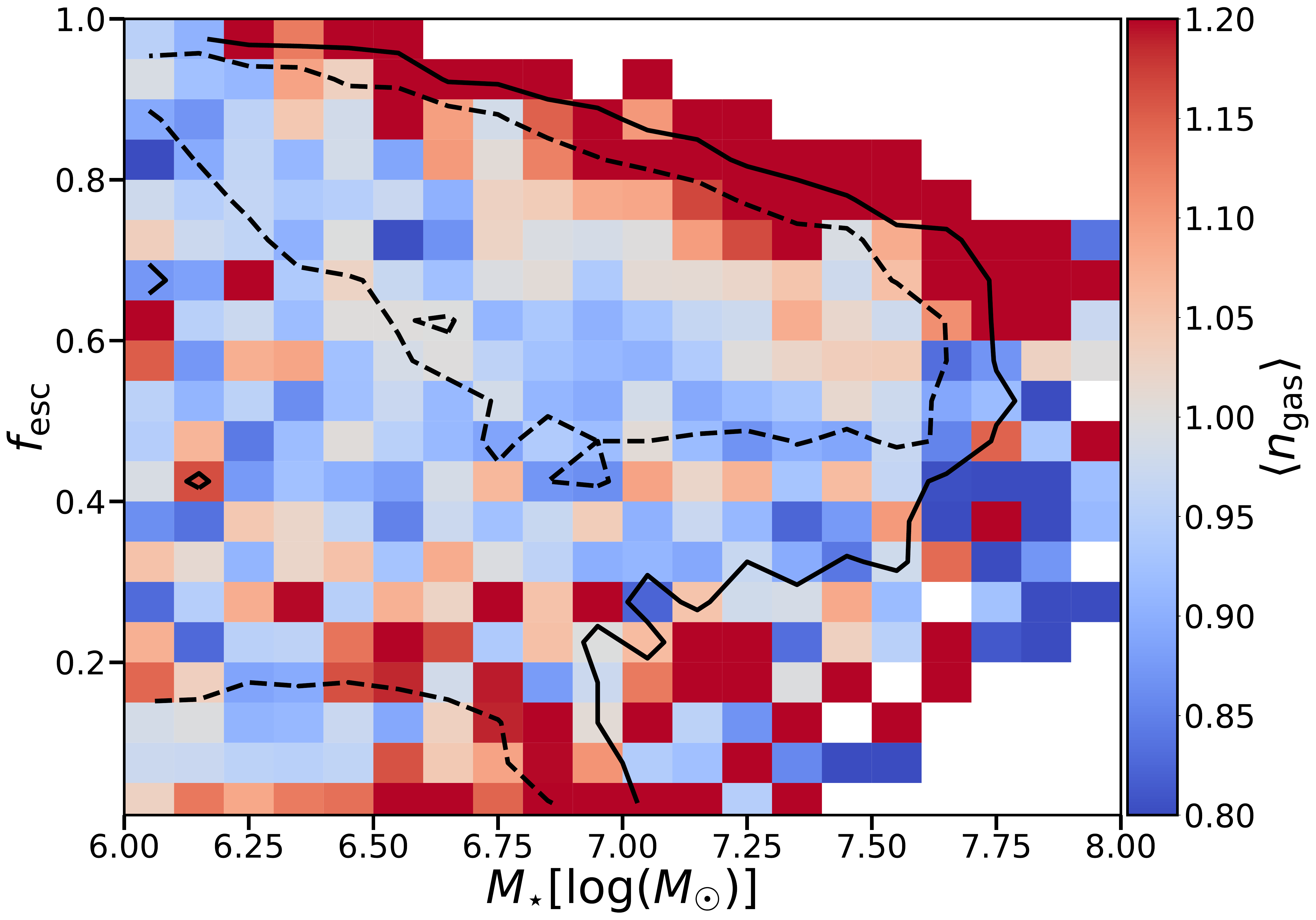}
    \includegraphics[width=0.43\textwidth]{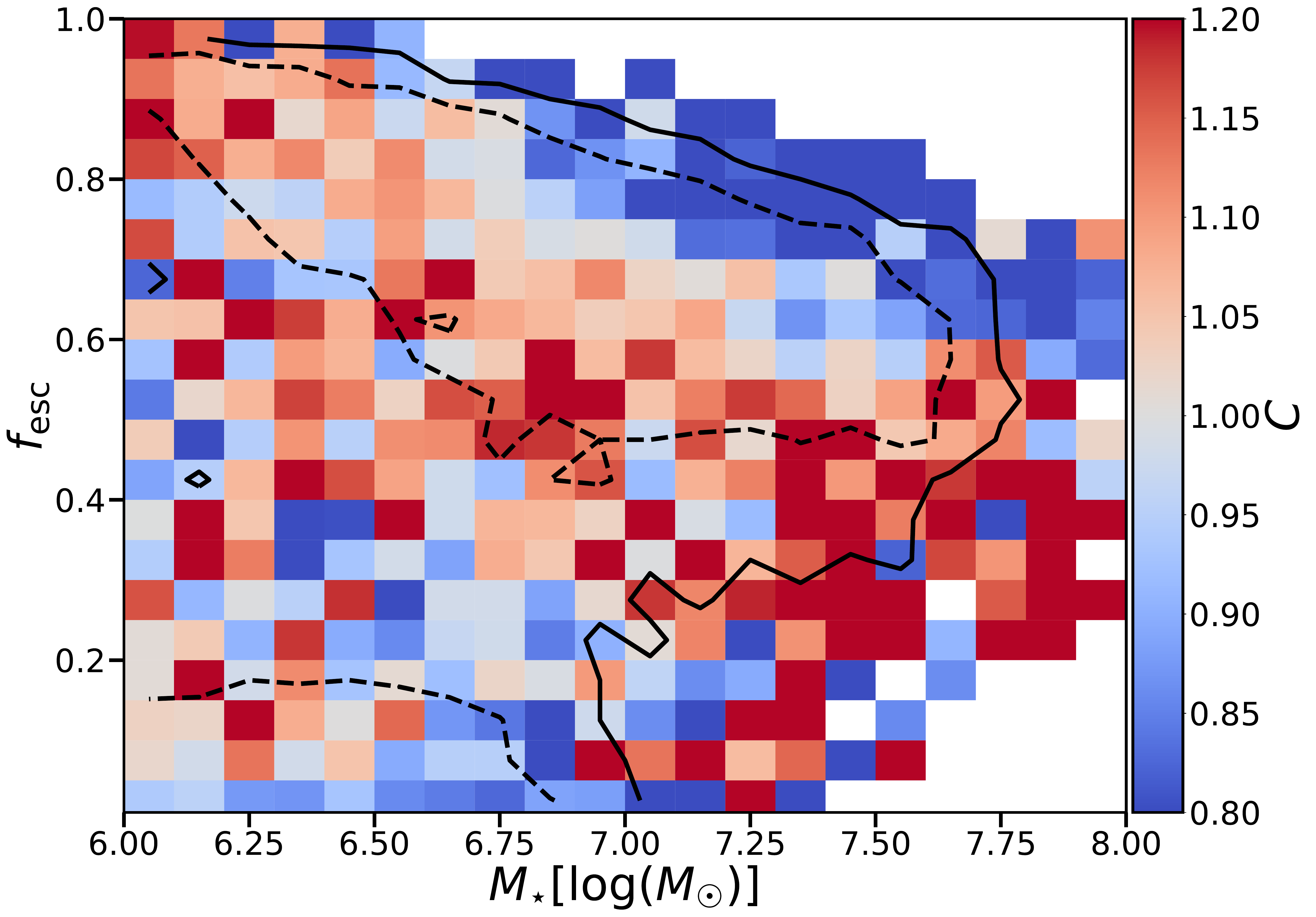}
    \includegraphics[width=0.43\textwidth]{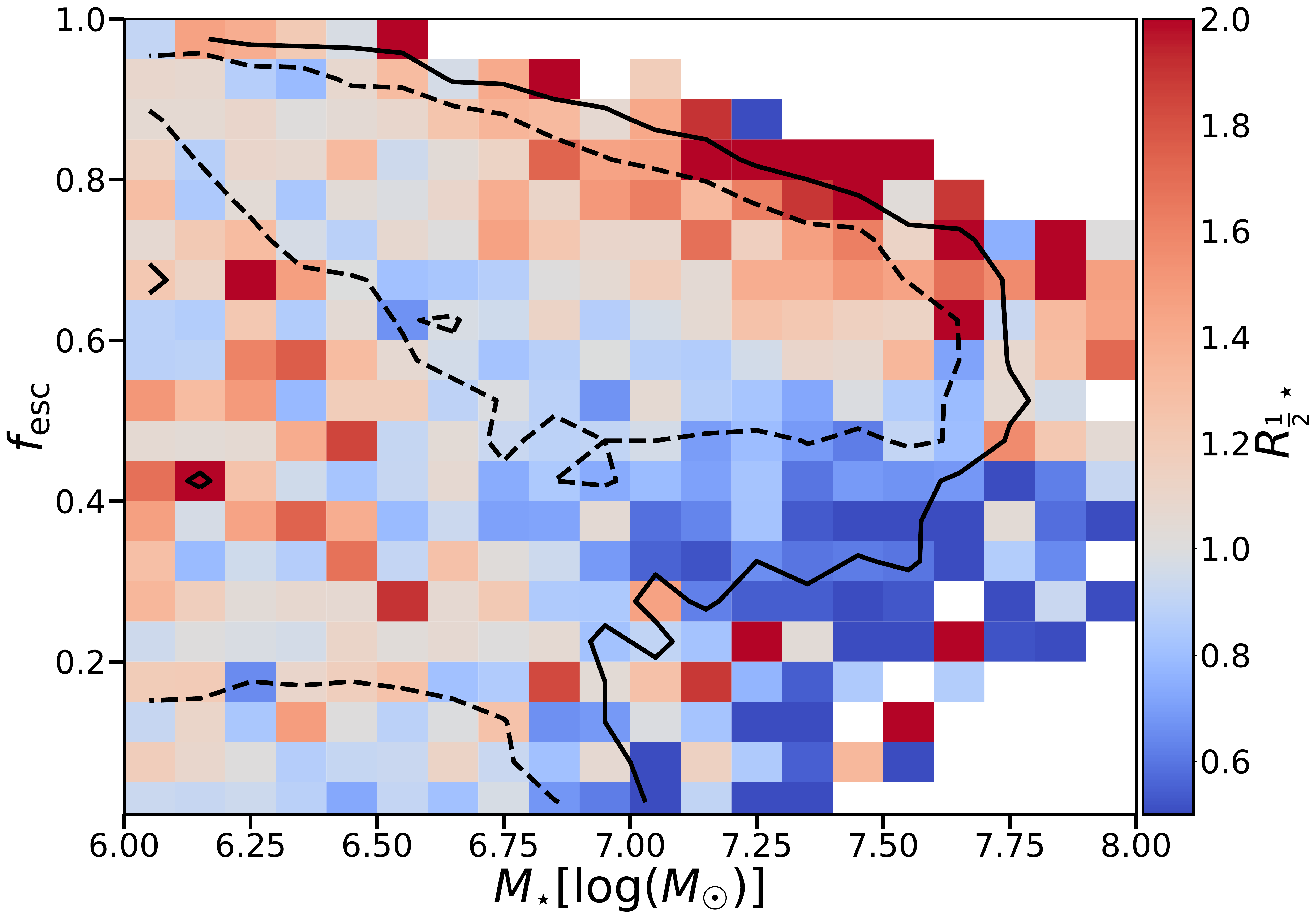}
    \includegraphics[width=0.43\textwidth]{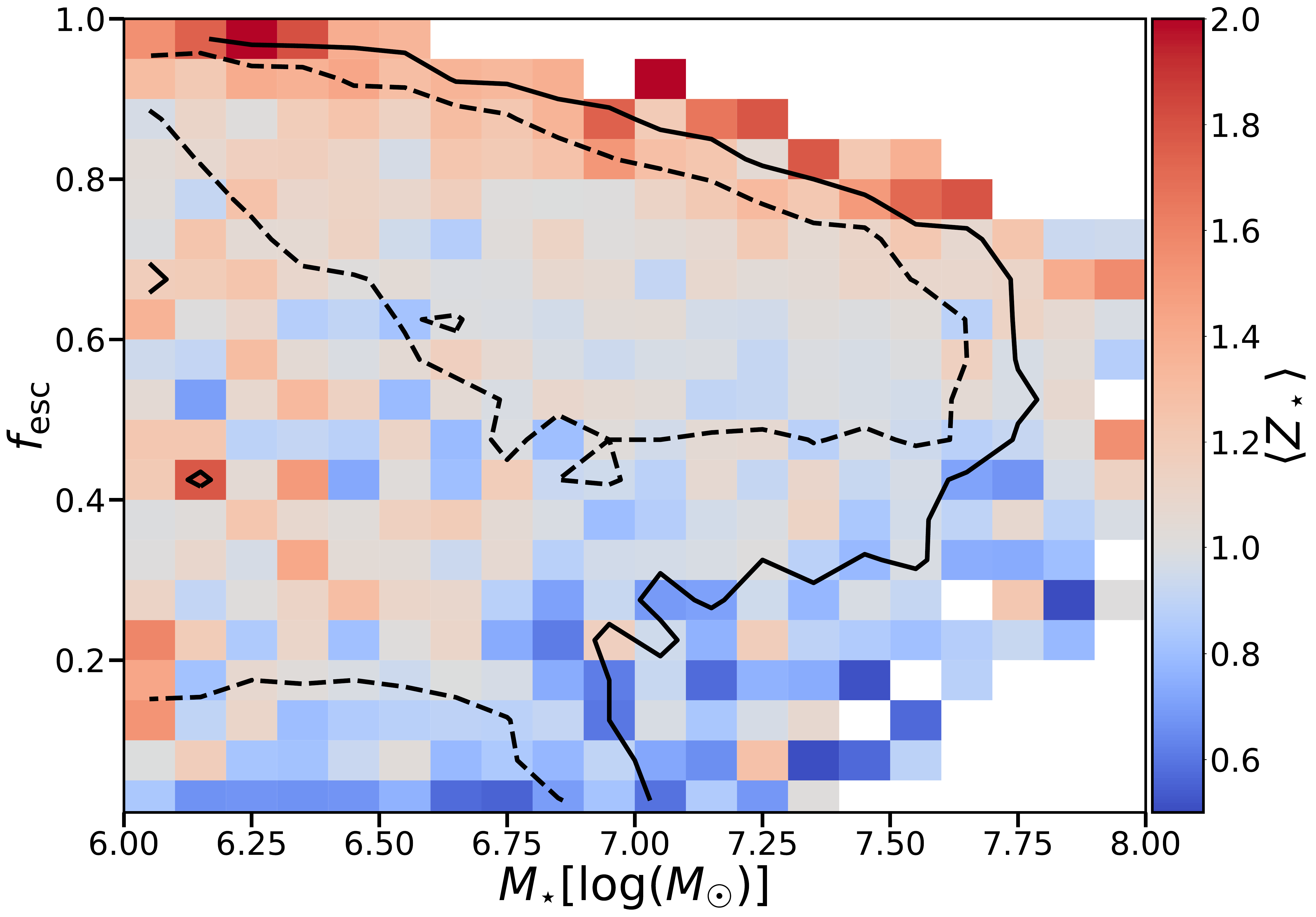}
    \caption{Ionizing escape fraction $f_\mathrm{esc}$ as a function of stellar mass for halos, combining $6 < z < 10$. In the eight panels, each colorbar and label refers to one of eight different physical properties. In each case, the color of every bin represents the median value of that quantity for all galaxies in that bin, normalized to the median value in that 0.1 dex stellar mass bin. That is, blue shows values lower than typical at that mass, while red shows values higher. From left to right, top to bottom, the properties are: gas surface density $\Sigma_{\rm gas, gal}$, galactic stellar surface density $\Sigma_{\star, \mathrm{gal}}$, halo's ionizing emissivity $Q_0$, virial radius of the halo $R_{\rm vir}$, average number density $\langle n_{\rm gas} \rangle$, halo clumping factor $C$, radius at which half the stellar emissivity of the halo is reached $R_{\rm 1/2, \star}$, and emissivity-weighted stellar metallicity $\langle Z_\star \rangle$. The solid and dashed lines represent the $1\sigma$ and $2\sigma$ contours of the number of galaxies in this plane. 
    }
    \label{fig:color1}
\end{figure*}

We calculate the mass surface density of stars $\Sigma_{\star, \mathrm{gal}}$ within the galaxy analogously as above, replacing gas mass with stellar mass. The top right panel of Figure \ref{fig:color1} is noisy, but broadly speaking we see a negative correlation for halos with $M_\star>10^{7.5}$M$_{\astrosun}$, and a positive/non-monotonic trend at lower masses. Indeed, the correlation of $f_\mathrm{esc}$ with stellar surface density at fixed mass is qualitatively similar, but inverted, with respect to the correlation with gas surface density. At intermediate masses, galaxies with both the lowest and highest escape fractions have relatively low $\Sigma_{\star, \mathrm{gal}}$, while those with average escape fractions have lower stellar surface densities.

We suggest that lower mass halos have less dense centers, such that young stellar particles are usually able to ionize their environment. Therefore, the escape fraction is mostly determined by the number of stars contributing to the ionization of the whole halo. The negative correlation for halos with a higher stellar mass, instead, may be due to the association of higher stellar surface density with higher gas density, so that stars are unable to fully ionize their immediate surrounding. As a result, a higher number of stars would not necessarily result in a stronger ionization of the halo.

\subsubsection*{Ionizing emissivity}

There is a clear and unambiguous positive correlation between escape fraction and the total ionizing emissivity of the halo, $Q_0$, particularly for lower mass halos. The large scatter of $f_\mathrm{esc}$ at $M_\star < 10^7$\,M$_\odot$ is due to variation in $Q_0$ - those galaxies with low to zero escape fraction have substantially lower total ionizing emissivities, by more than a factor of four (where we saturate the relative color range). At these low masses, the escape fraction is often dominated by a few young stellar populations, and the location of these sources with respect to the underlying gas distribution in the galaxy and halo is key. Depending on their position within the halo, these can either fully ionize the surrounding medium, or barely ionize it at all.

At the high-mass end of our study, $10^{7.5} < M_\star / \rm{M}_\odot < 10^8$, the correlation of escape fraction with total ionizing emissivity has the same sign, although it is weaker. In particular, high $f_\mathrm{esc}$ galaxies are always those with larger $Q_0$, as expected.

\subsubsection*{Halo virial radius}

At fixed stellar mass, galaxies can reside in dark matter halos of varying total mass, and thus size. To explore this, we examine the correlation with the virial radius of the parent halo. As with gas surface mass density, we find that galaxies with typical $f_\mathrm{esc}$ values sit in halos with typical (median) virial radii, whereas outliers in escape fraction, either high or low, tend to sit in smaller (more compact) halos. This trend is most visible at intermediate stellar masses, and is negligible at both the low-mass and high-mass ends of our sample.

Together with the observed correlation with gas surface density, this result suggests that more compact halos hosting less compact galaxies tend to have the highest escape fractions.

\subsubsection*{Average gas density}

The propagation of ionizing radiation is strongly impacted by the surrounding gas density structure. To zeroth order, we therefore examine the connection between escape fraction and the average gas number density of the halo $\langle n_\mathrm{gas} \rangle$. We again observe a characteristic non-monontonic correlation whereby galaxies at typical values of $f_\mathrm{esc} \sim 0.6$ have the lowest average gas densities, whereas outliers both high and low reside in denser halos.

This effect is qualitatively similar to that observed for stellar surface density, but even more clear and pronounced. If the overall gas density in the halo is higher than is typical at a given mass, then the escape fraction can be driven either low (for low total emissivity), or high (for high total emissivity); in both cases the stellar source distribution is also (overly) compact.

\subsubsection*{Halo gas clumping factor}

The gas density field encountered by propagating photons is not smooth. We therefore define and measure the gas clumping factor as
\begin{equation}
    C = \langle n_\mathrm{gas}^2 \rangle_V / \langle n_\mathrm{gas} \rangle^2_V,
\end{equation}
where the average of gas number density is taken over the volume $V$ of the halo. Overall, we see that escape fraction negatively correlates with clumping. Since the recombination rate scales as density squared, a stronger clumping facilitates more recombinations, making it more difficult for photons to create an ionized escape channel. On the other hand, stronger clumping also means that the gas distribution is more localized and less homogeneously distributed, which could lead to more underdense regions and facilitate the creation of ionized channels, increasing the escape fraction.

However, the latter effect seems to be subdominant. This roughly agrees with \cite{hattab19} and FiBY, which found a strong correlation between escape fraction and clumping for only very low values of clumping, while we see that relatively small variations in $C$ significantly change $f_{\rm esc}$. Higher resolution simulations could potentially lead to a different result, as clumping on smaller scales leads to the formation of more clustered star formation and so coherent emission of ionizing radiation \citep{ciardi05}.

\subsubsection*{Stellar half emissivity radius}

The galaxy stellar `size', emissivity weighted, is calculated as the stellar half emissivity radius (lower left panel). Above a stellar mass of $M_\star \gtrsim 10^{6.5}\,$M$_\odot$ we find a clear, positive correlation, such that galaxies with more extended stellar source distributions have higher escape fractions. This connection is reassuring and as expected: when stellar populations are less centrally concentrated, they are less surrounded by the densest gas at the center of the halo, facilitating the ionization of the halo gas and higher $f_\mathrm{esc}$.

\subsubsection*{Emissivity weighted stellar metallicity}

Finally, we find a clear positive correlation between the emissivity-weighted stellar metallicity and the escape fraction (lower right panel). As higher metallicity stars have longer lifetimes, this increases the time where such stellar populations have the chance to leave their initial birth clouds while still young enough to produce significant amounts of ionizing radiation. 

In addition to the quantities explored here, we also considered halo metallicity, the gas fraction of the halo, total halo mass itself, the galaxy SFR as well as the emissivity weighted age of the stellar populations (not shown). In these cases any trends with escape fraction are weak and tentative at best. Although a correlation with SFR would be expected, we suspect that the instantaneous values of SFR we use, which oscillate on short timescales, are not sufficiently connected to the ionization state of the gas. Short time-scale variation of SFR, over different time periods, is expected to be connected to highly time variable $f_\mathrm{esc}$ \citep{rosdahl22}.

\subsection{Impact of the local escape fraction}
\label{sec:fescloc}

In all numerical hydrodynamical simulations with finite resolution, small-scale gas structure will be resolved down to some scale, but unresolved for smaller scales. Although the resolution of TNG50 is comparably high, achieving gas cell sizes of order a few physical parcsecs at these redshifts, it does not well resolve parsec-scale ISM structure. Gas at these scales would realistically impact the small-scale escape of ionizing radiation \citep{dale13, howard18, kim19, kimm19,he2020}. As a result, some simulations adopt a sub-grid escape fraction, also referred to as a local escape fraction, or unresolved escape fraction. This value accounts for photon absorption on subgrid scale, i.e. the fraction of ionizing photons that escape their cell of origin. 

In many RHD simulations of reionization, this $f_\mathrm{esc, loc}$ parameter is taken to be unity, which assumes that there is no unresolved gas density structure of relevance. The validity of such a choice depends sensitively on the numerical resolution of the underlying simulation. For our fiducial results, we have also adopted $f_\mathrm{esc, loc} = 1$, as the simplest possible choice. We now test the validity and impact of this assumption for the specific case of TNG50.

\begin{figure}
    \centering
    \includegraphics[width=0.45\textwidth]{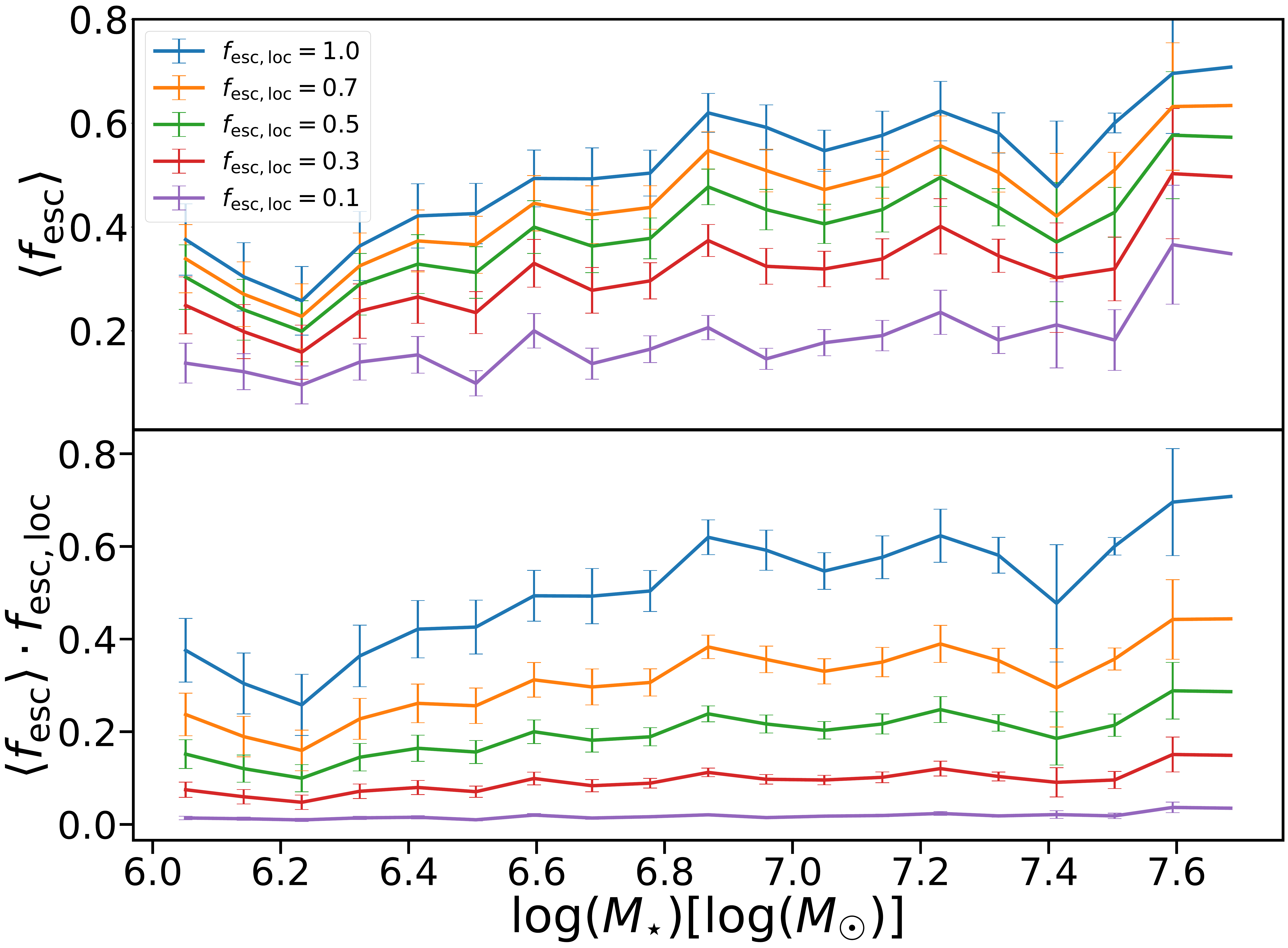}
    \caption{{\it Top:} Average halo escape fraction of ionizing photons as a function of stellar mass for different values of local escape fraction, $f_\mathrm{esc, loc}$, as indicated by the colors. The error bars show the standard deviation on the mean in each mass bin, and here we combine halos at all three examined redshifts. \textit{Bottom:} Average total escape fraction, obtained multiplying the halo escape fraction from the top panel by the corresponding local escape fraction. Measured values of $f_\mathrm{esc}$ depend sensitively, and non-linearly, on the value of the local escape fraction assumed.
    }
    \label{fig:unresolved}
\end{figure}

To do so, we have repeated our RT calculations adopting different values of $f_\mathrm{esc, loc}$, and Figure \ref{fig:unresolved} shows the result. In the top panel we show the average escape fraction obtained with different assumptions for $f_\mathrm{esc,loc}$, from a value of 10\% to the fiducial case of 100\%. As expected, a higher $f_\mathrm{esc,loc}$ results in a higher fraction of photons which can escape the halo, as the larger production rate of ionizing photons ionizes more of the halo gas. There is an increase of $5-10\%$ in the escape fraction for every $20\%$ increase in the local escape fraction, with the exception of $f_\mathrm{esc,loc}=0.1$ where the difference is in the range of $10-15\%$. This effect is more strongly seen for more massive halos, while for halos with a stellar mass of $M_\star<10^{6.5}$~M$_{\astrosun}$ the variation is smaller. For the smallest local escape fraction of $f_{\mathrm{esc,loc}}=0.1$ there is almost no evolution in the halo escape fraction with stellar mass.

There is a clear, non-linear dependence of $f_\mathrm{esc}$ on the value of the local escape fraction, and thus the global escape fraction cannot be simply obtained by re-scaling. Instead, separate RT simulations must be run for different values of $f_\mathrm{esc,loc}$.

In any of the cases with $f_\mathrm{esc,loc} < 1 $, the halo escape fraction, i.e. the ratio of emitted photons to halo escaping photons, is given by the multiplication of the local escape fraction parameter with the measured fraction, as described above. The bottom panel of Figure \ref{fig:unresolved} shows these values, appropriate for comparison with observationally inferred escape fractions or for inclusion into large-scale models where the entire process of photon escape is unresolved.

\subsection{Ionizing photon emissivity} \label{sec:emissivity}

\begin{figure*}
    \centering
    \includegraphics[width=0.45\textwidth]{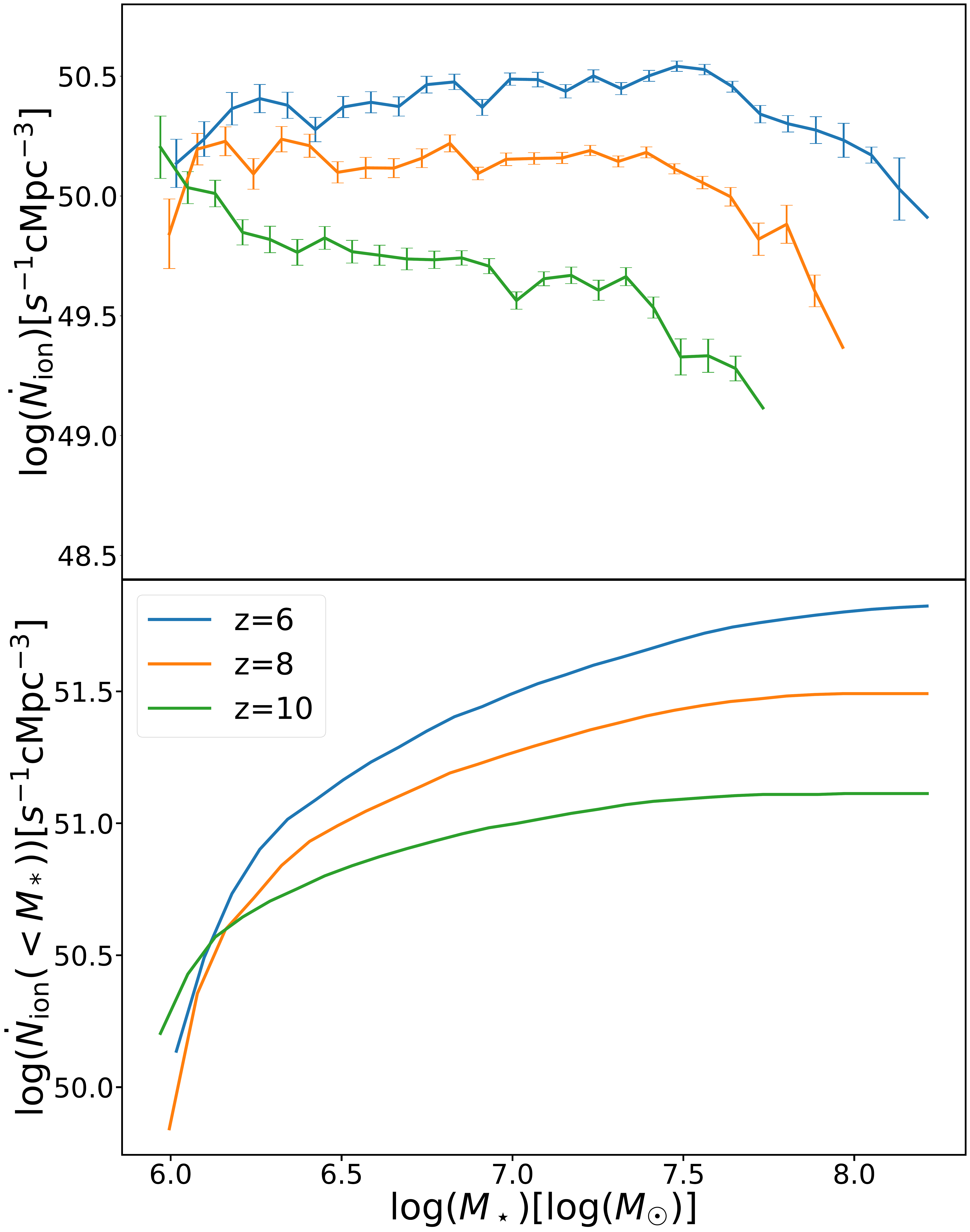}
    \includegraphics[width=0.45\textwidth]{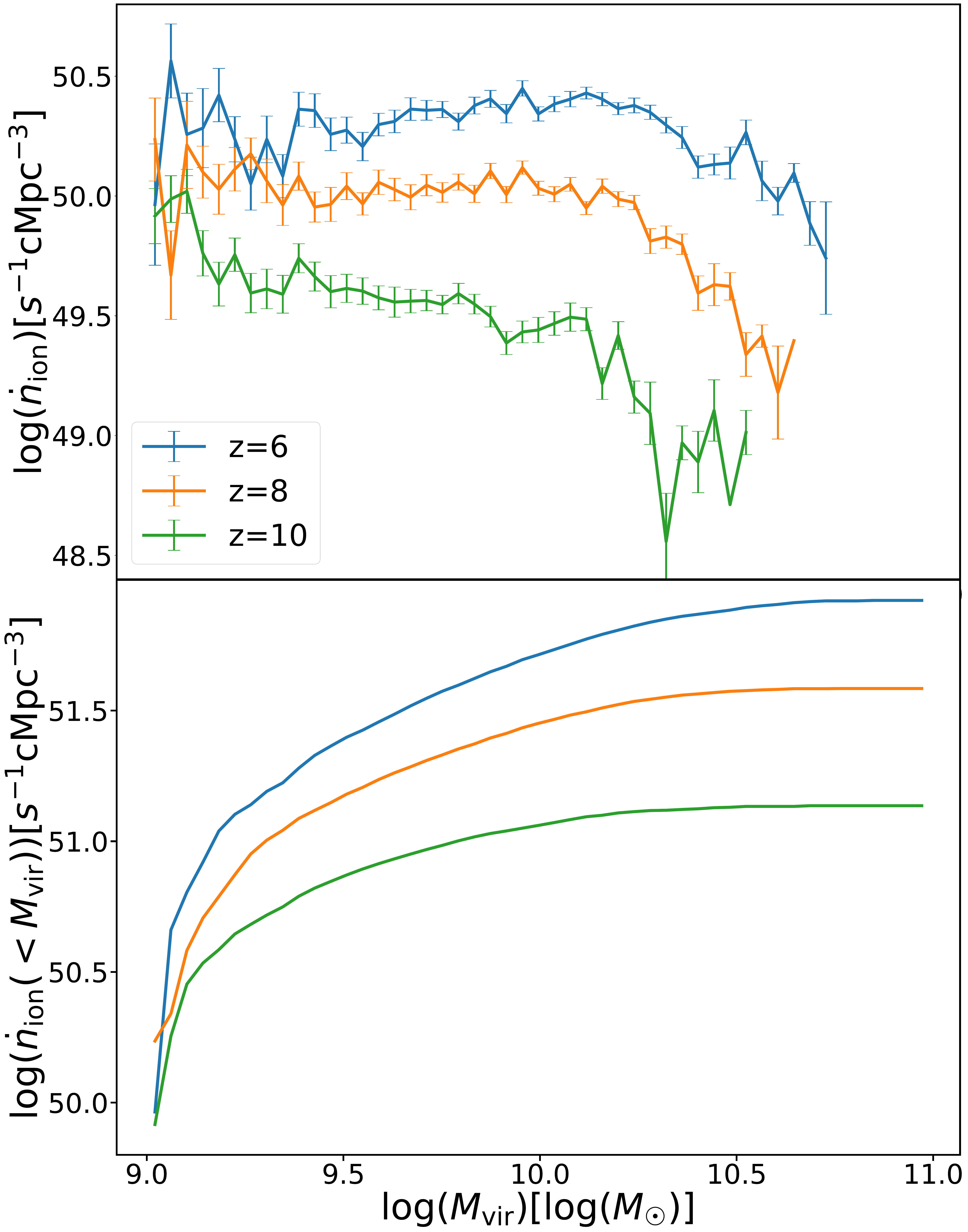}
    \caption{\textit{Left:} Global ionizing photon emissivity (top panel) and cumulative emissivity (bottom) as a function of the stellar mass for halos at $z=6$ (blue lines), $z=8$ (orange) and $z=10$ (green) as predicted by our TNG50+CRASH RT simulations. The error bars in the top panel represent the propagated error of the escape fraction. \textit{Right:} The same quantities, shown as a function of total halo mass instead of galaxy stellar mass. Galaxies with a stellar mass of $M_\star<10^{7.5}$M$_{\astrosun}$ (halos with $M_\mathrm{vir}<10^{10}$M$_{\astrosun}$) contribute the bulk of the ionizing photons at all redshifts.
    \label{fig:emisdensity_star}
    }
\end{figure*}

In Figure \ref{fig:emisdensity_star} we investigate the contribution of halos of different mass to the total production of ionizing photons. To calculate the emissivity, $\dot{N}_{\rm ion}$, we weight the average number of ionizing photons escaping from objects in a given mass bin with the density of halos within the same bin, obtained from the halo abundance in the TNG50 catalog. Throughout this study we use the results of our fiducial run, i.e. $f_\mathrm{esc,loc}=1$. 

Halos with a stellar mass of $M_\star<10^{7.5}$M$_{\astrosun}$ ($M_\mathrm{vir}<10^{10}$M$_{\astrosun}$) contribute the largest number of ionizing photons at all redshifts. While the maximum of the emissivity at $z=10$ corresponds to $M_\star<10^6$M$_{\astrosun}$ ($M_\mathrm{vir}<10^{9}$M$_{\astrosun}$), such halos are not included in this study as they are not well resolved. Therefore, the budget of ionizing photons presented here could be seen as a lower limit at $z \gtrsim 10$. Our upper mass cutoff, instead, excludes the heaviest halos. However, as the emissivities decrease significantly at the higher mass end, visible in the flattening of the global emissivity curves, we expect these halos to play a negligible role in the photon budget.

At $z=8$ the emissivity reaches its maximum at $M_\star \approx 10^{6.5}$M$_{\astrosun}$ ($M_\mathrm{vir} \approx 10^{9.25}$M$_{\astrosun}$), while at $z=6$ we see a peak at $M_\star \approx 10^{7.5}$M$_{\astrosun}$ ($M_\mathrm{vir} \approx 10^{10}$M$_{\astrosun}$), after which $\dot{N}_{\rm ion}$ decreases significantly. At these redshifts, though, the curves are almost flat over about an order of magnitude in mass.
At $M_\star \sim 10^6$M$_{\astrosun}$ ($M_\mathrm{vir} \sim 10^9$M$_{\astrosun}$) the emissivity at different redshifts does not vary significantly, while it declines with increasing mass for all redshifts. Since the escape fraction is not strongly dependent on redshift, the difference in emissivity can be mostly attributed to the lower abundance of massive halos at higher redshifts.

In the bottom panels of Figure \ref{fig:emisdensity_star} we plot the cumulative emissivity. As before, we observe a rapid rise in the range $M_\star \sim 10^6-10^7$M$_{\astrosun}$ ($M_\mathrm{vir} \sim 10^9-10^{10}$M$_{\astrosun}$) for all redshifts, while at higher masses the emissivity begins to saturate. Halos with a stellar mass of $M_\star \gtrsim 10^7$M$_{\astrosun}$ ($M_{\mathrm{vir}} \gtrsim 10^{10}$M$_{\astrosun}$) do not significantly contribute to the budget of ionizing radiation at high redshifts.

\begin{figure}
    \centering
    \includegraphics[width=0.45\textwidth]{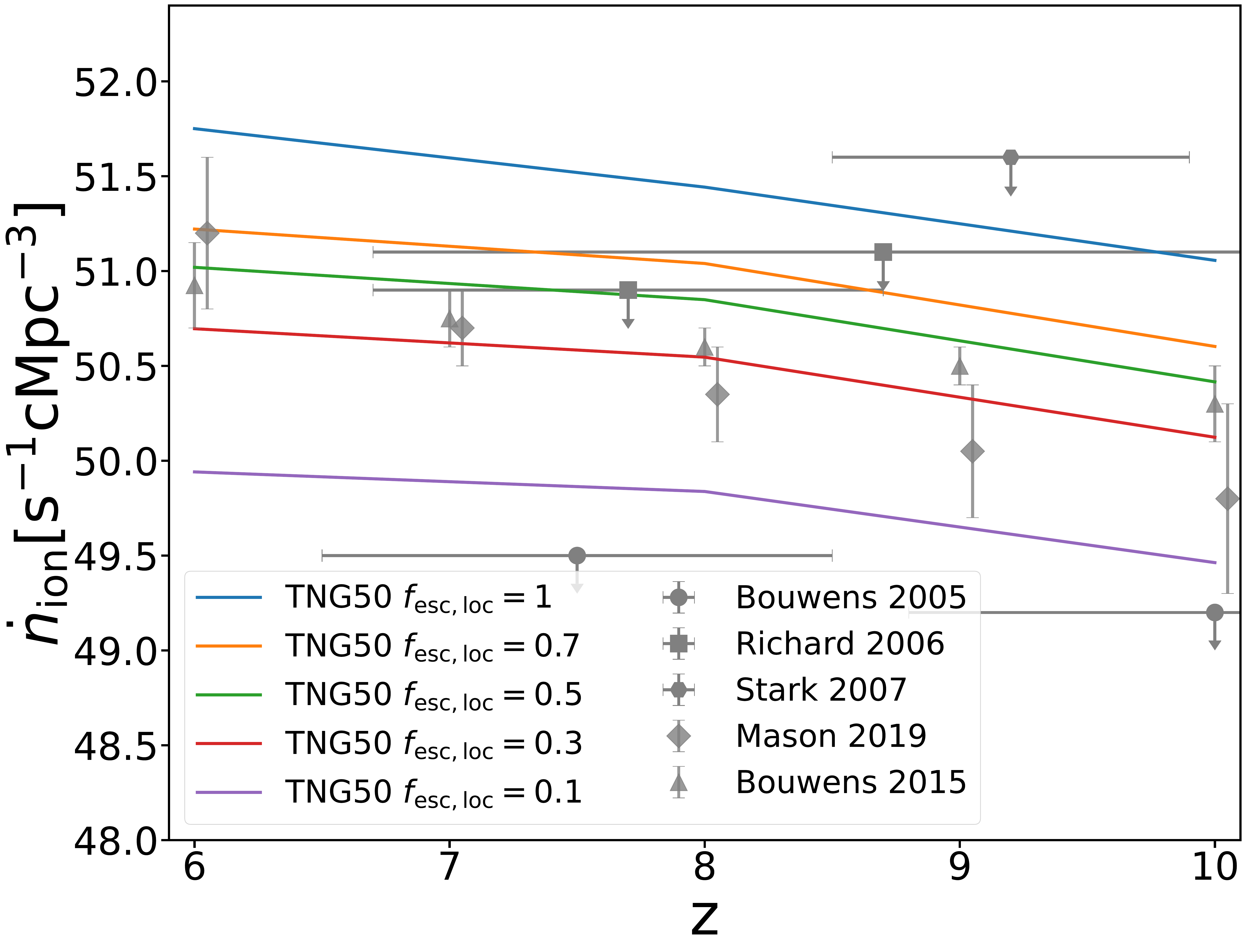}
    \caption{Ionizing photon emissivity predicted by the TNG50 and CRASH simulations as a function of redshift for various settings of the local escape fraction (indicated by lines of different color). Symbols refer to data, as indicated in the labels. 
    }
    \label{fig:emisdensity_z}
\end{figure}

In Figure \ref{fig:emisdensity_z} we compare the emissivities obtained with our model to observational constrains from \cite{bouwens15} and \cite{mason19}, in addition to the results of \cite{stark07}, \cite{richard06}, and \cite{bouwens05} as presented in \cite{bolton07} with the assumption of a constant $f_\mathrm{esc}=0.2$. This is a comparison at face value, i.e. we have not replicated the selection functions of the observed data nor the measurements via e.g. forward modeling.

We find that while our prediction with $f_\mathrm{esc,loc}=1.0$ lies above most observational constraints, and is thus at face-value disfavored, adopting $f_\mathrm{esc,loc}=0.1$ tends to underestimate the data. Values in the range $f_\mathrm{esc,loc}=0.3-0.7$ are in the best agreement with observational constraints -- for the specific case of TNG50, i.e. the realization of the TNG galaxy formation model at the numerical resolution of TNG50. We further note that we have adopted only globally constant values of $f_\mathrm{esc,loc}$, i.e. the local escape fraction parameter does not depend on redshift nor mass. In reality, the dependence is likely more complex.

\subsection{Single source escape fractions}

\begin{figure*}
    \centering
    \includegraphics[width=0.49\textwidth]{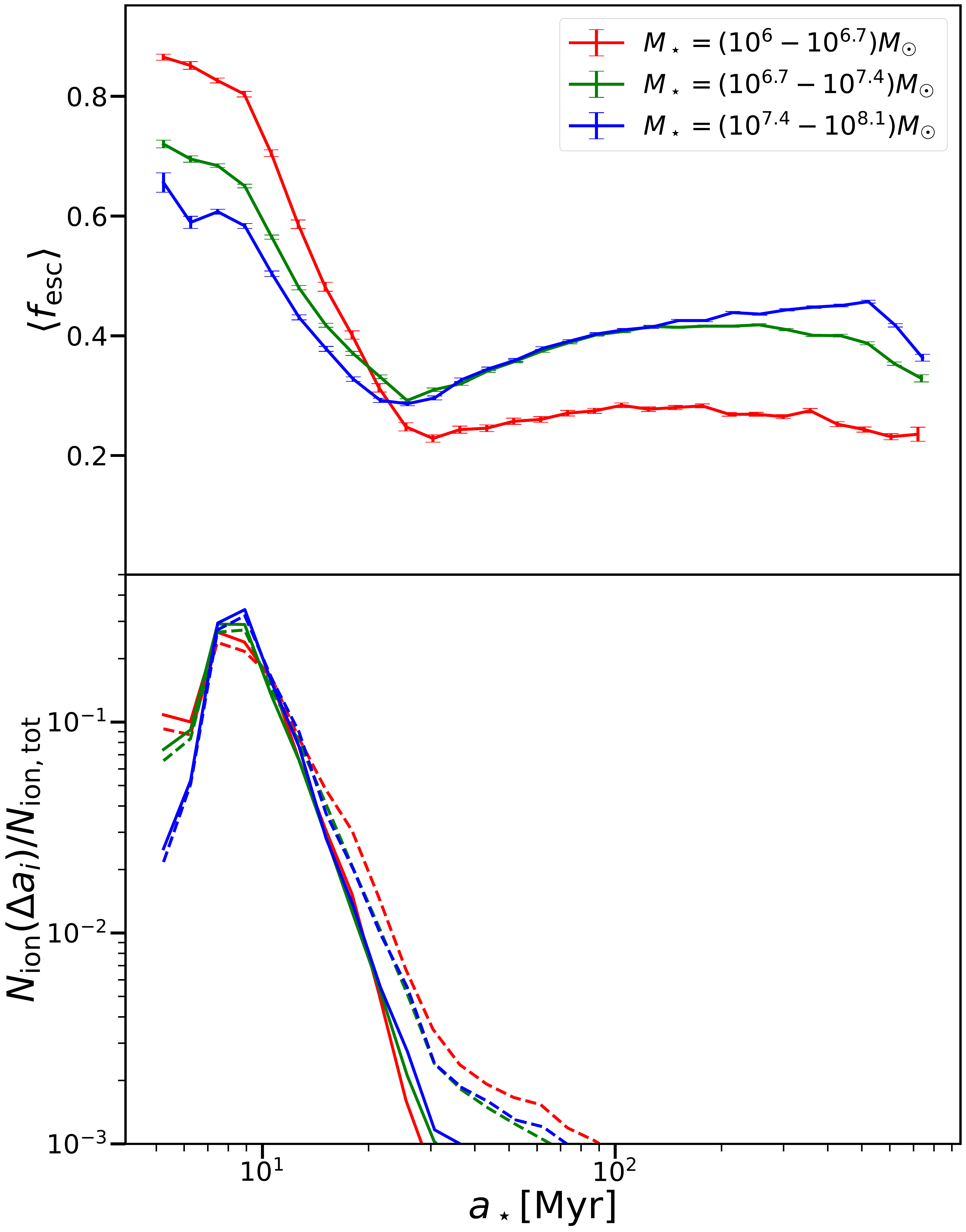}
    \includegraphics[width=0.49\textwidth]{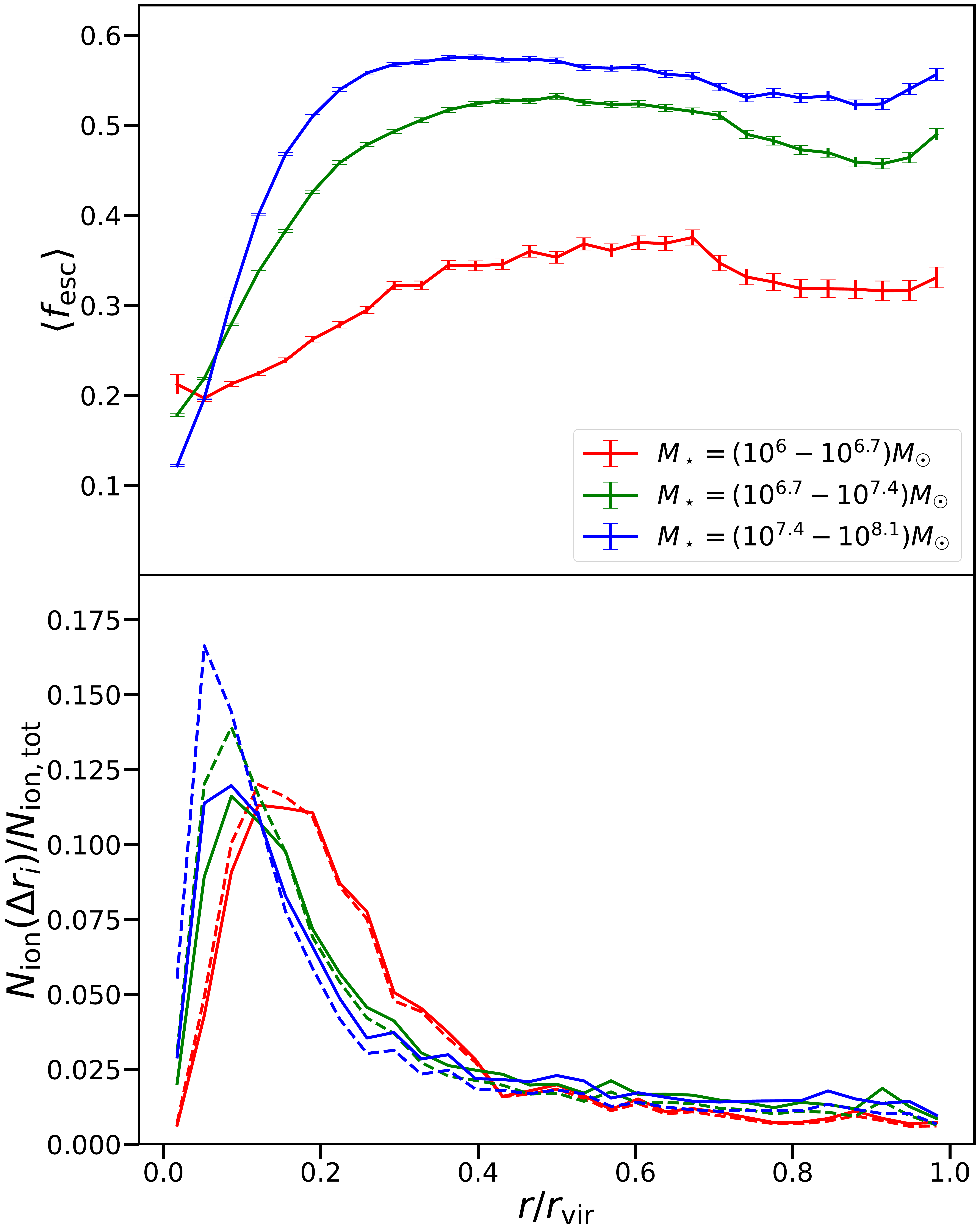}
    \caption{\textit{Top panels:} Average halo escape fraction of individual stellar particles as predicted by the TNG50 and CRASH simulations, as a function of their age (left) and distance from the center of the halo (right) in three bins of stellar mass, indicated with the different colors. The errorbars represent the errors of the mean values. \textit{Bottom panels:} Relative contribution of stars in a given age (left) and distance from the center (right) bin to the total ionizing photon budget. The solid and dashed lines denote the contribution to the escaped and emitted ionizing photons, respectively. Intermediate age stars born $\sim\,$tens of Myr ago, especially those not in the densest central region of the galaxy, have the largest escape fractions, while younger $\sim\,10$\,Myr old stars on the outskirts of galaxies have the largest fractional contribution to the total ionizing photon budget.}
    \label{fig:fesc_r}
\end{figure*}

We next examine the ability of ionizing photons to escape from individual stellar populations, in order to explore the stars that contribute, or not, to the budget of ionizing radiation escaping the halos.

The top panels of Figure \ref{fig:fesc_r} show the escape fraction from individual star particles as a function of their age (left column) and distance from halo center (right column). We separate halos into three different stellar mass bins, namely, $M_\star=10^{6.0-6.7}$M$_{\astrosun}$, $10^{7.0-7.4}$M$_{\astrosun}$ and $10^{8.0-8.1}$M$_{\astrosun}$ (red, green, and blue lines).

The average escape fraction as a function of stellar age (top left panel) decreases with increasing age until it reaches a minimum at $\sim 20-30 \%$ for ages of $\sim\,20-30$\,Myr, before increasing slightly for older populations. As we have previously seen, young stars produce abundant ionizing photons, which are able to efficiently ionize the dense gas surrounding their formation sites and eventually escape. On the other hand, older stars are less efficient at forming ionized channels in the gas, resulting in a lower escape fraction. At the same time, older stars are less clustered within clouds of the densest and most neutral gas, enabling their relatively low production of ionizing photons to induce high escape fractions. 

With respect to trends with galaxy mass, the youngest stellar particles tend to have the highest escape fractions for smaller halos, while the trend is reversed for the oldest stellar particles. We speculate that this behaviour is driven by the high central gas densities in massive halos, hindering the escape of the radiation from young stars.

The average escape fraction as a function of distance (top right panel) shows that close to the halo center the average escape fraction is smallest ($\approx 10-20\%$), as stellar particles located there are embedded in the denser gas. As we move further away from the center, the gas becomes less dense, resulting in higher escape fractions. Any sources which are located outside the central galaxy, i.e. in the circumgalactic medium, have large $f_\mathrm{esc}$, although the total luminosity of such sources is low. At large distances, the escape fraction plateaus to a constant value of $\approx 50\%$ and $30\%$ for heavier and lighter halos, respectively. Here the decreasing gas density reaches a balance with the decreasing density of stellar sources.

The evolution of the escape fraction with radius is qualitatively similar for all mass bins. However, the effects of the absorption in the central regions is strongest for the most massive halos, and smallest for the least massive. Note that although the spread of the escape fraction of individual stellar particles is large, our sample size is sufficient to identify differences in the mean trends.

The bottom panels of Figure \ref{fig:fesc_r} show the fractional contribution of stars to the total budget of escaping and emitted photons, again as a function of their age (left) and halo-centric distance (right). The contribution to the escaping photons (solid lines) peaks at a stellar age of $a_\star \approx 10$Myr. The behaviour for $a_\star \gtrsim 10$Myr is very similar in all mass bins, while for younger stars the relative contribution is more significant for low-mass halos. Through a comparison to the emitted photons (dashed lines), a difference is seen in the two contributions only for $a_\star \gtrsim 20$Myr. In short, we find that while younger stars contribute to the emitted and emerging photon budget more or less equally, older stars contribute substantially more to the former than to the latter. This is due to stars whose emissivity is not enough to ionize the surrounding dense gas, corresponding to the minimum in the escape fraction observed in the top-left panel.\footnote{Note, however, that these differences appear for stellar ages which contribute little to the total ionizing photon budget. They are, therefore, not of great importance for the resulting $f_\mathrm{esc}$. Even older stars have such a small contribution to the emission budget that they lie outside the range depicted.}

As a function of distance (bottom right panel), the fractional contribution for both emitted and escaping photons rises quickly at small radii, reaching a peak at $\sim\,0.1\,R_{\rm vir}$, before slowly declining at the very center. This peak shifts outwards and broadens with decreasing stellar mass. Emitted versus escaped photons show similar behavior in the smallest stellar mass bin. For higher masses, escaped photons exhibit a flatter distribution than their emitted counterparts. This reflects the fact that in larger halos ionizing photons emitted near the center are more easily absorbed by the surrounding gas.

\subsection{Frequency dependence and the escaping spectra} \label{sec:spectra}

\begin{figure}
    \centering
    \includegraphics[width=0.47\textwidth]{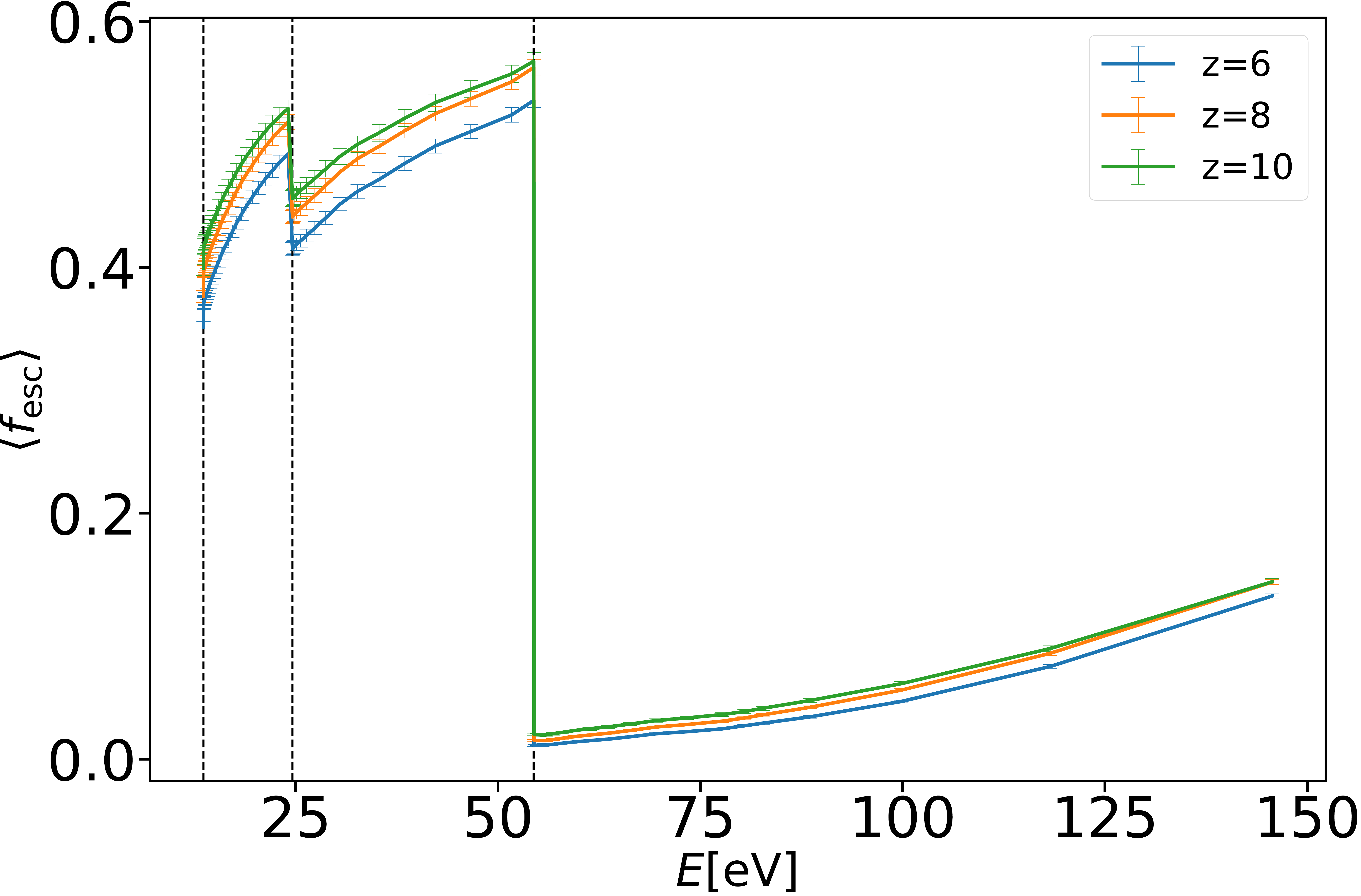}
    \caption{Average halo escape fraction as a function of photon energy for three different redshifts as predicted by the TNG50+CRASH RT simulations. The vertical dotted lines denote, from left to right, the ionization energies of HI, HeI and HeII. The error-bars correspond to the error of the mean value in a given energy bin. The results refer to our fiducial model with $f_\mathrm{esc,loc}=1$. Escape fractions at high energies $E > 54.4$eV are substantially suppressed with respect to lower photon energies.}
    \label{fig:freq_fesc}
\end{figure}

Moving beyond a single (gray) characteristication of the escape fraction, we now take advantage of our multi-frequency approach to study how the escape fraction depends on photon frequency, as well as the impact on the emergent, versus intrinsic, spectra.

Figure \ref{fig:freq_fesc} shows the average escape fraction as a function of photon energy. We examine the halos in our fiducial configuration, i.e. $f_\mathrm{esc,loc}=1$. Due to the frequency dependence of the cross section, the escape fraction has a local minimum at each of the ionization thresholds of hydrogen and helium. Rightwards of each such feature, $f_\mathrm{esc}$ then rises for higher energies, as gas becomes increasingly transparent to these frequencies.

The escape fraction is also dramatically reduced at energies just above the HeII ionization threshold because halos have abundant HeII -- see the examples in Figure \ref{fig:halo_image}. This is largely due to the paucity of photons emitted at these frequencies, and this reduction acts in addition to the cross section effect.

We note only a slight evolution in the frequency-dependent escape fraction with redshift, where halos at $z=10$ have an approximately $5\%$ higher $f_\mathrm{esc}$ than halos at $z=6$. Interestingly, the difference between the escape fraction at $z=8$ and $z=10$ is much smaller than between $z=8$ and $z=6$, suggesting a faster evolution of the escape fraction with redshift towards the end of reionization.

\begin{figure}
    \centering
    \includegraphics[width=0.45\textwidth]{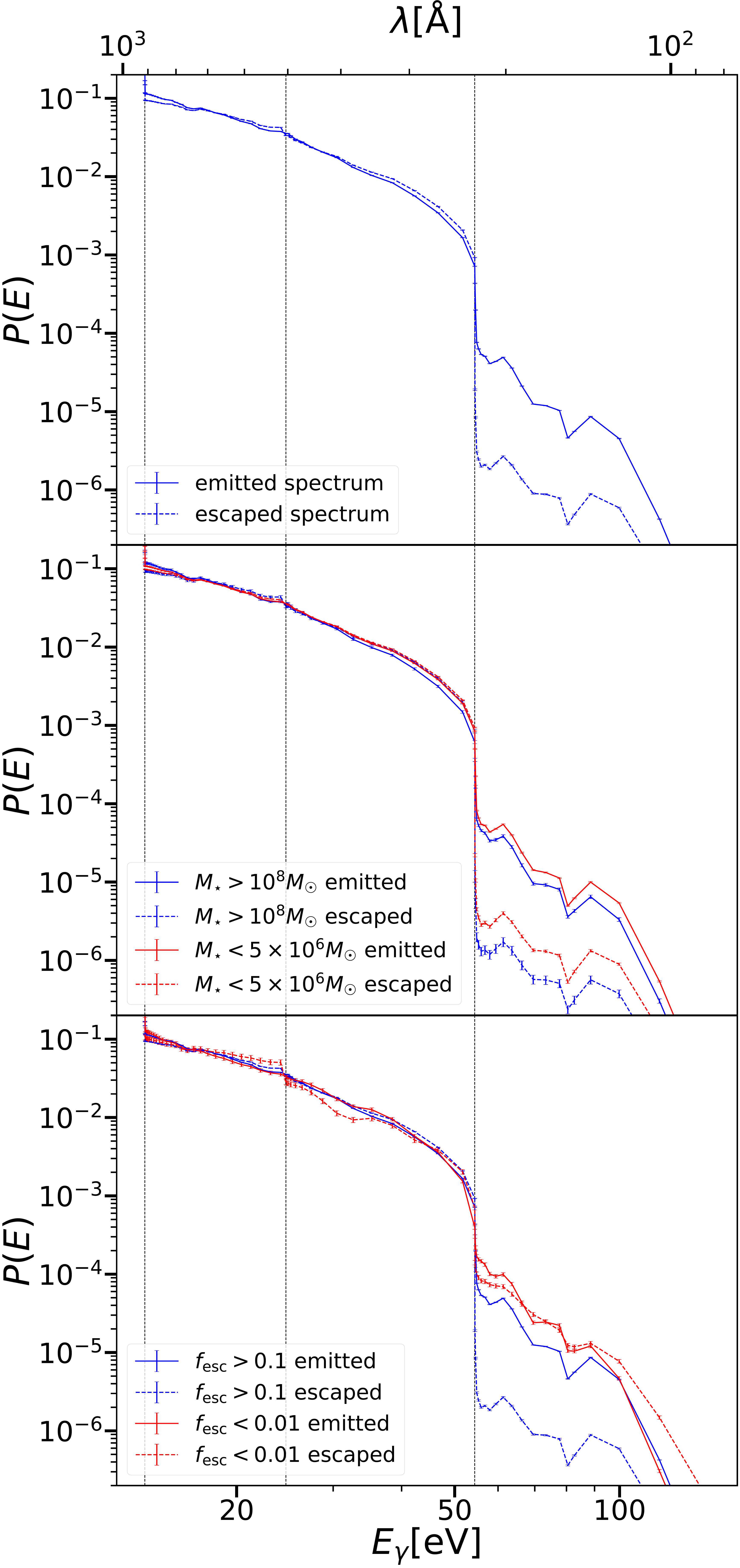}
    \caption{Normalized spectra of the emitted (solid lines) and escaping (dashed) ionizing photons for halos at all redshifts as predicted by the TNG50+CRASH RT simulations. The spectra have been calculated combining all halos (top panel); considering separately halos with stellar masses higher than $10^8$\,M$_{\astrosun}$ and lower than $10^6$\,M$_{\astrosun}$ (middle panel); and considering separately halos with an escape fraction higher than 10\% and lower than 0.1\% (bottom panel). In each case, the vertical dotted lines denote, from left to right, the ionization energies of HI, HeI and HeII.}
    \label{fig:spectrum}
\end{figure}

Figure \ref{fig:spectrum} shows how the actual spectra are modulated by escape. We plot the normalized sum of spectra\footnote{As a result, the features of individual spectra are averaged out.}, i.e. the volume enclosed by each curve is the same. Most of the escaping radiation has energies below the HeII ionization at $54.4\mathrm{eV}$. 

The top panel contrasts emitted versus escaping ionizing photons, and we see that the escaped spectrum has a lower amplitude at, or directly after, the ionization thresholds of hydrogen and helium. This effect is particularly visible after the HeII line at $54.4$eV, reflecting the behaviour observed in Figure \ref{fig:freq_fesc}.

\begin{figure}
    \centering
    \includegraphics[width=0.45\textwidth]{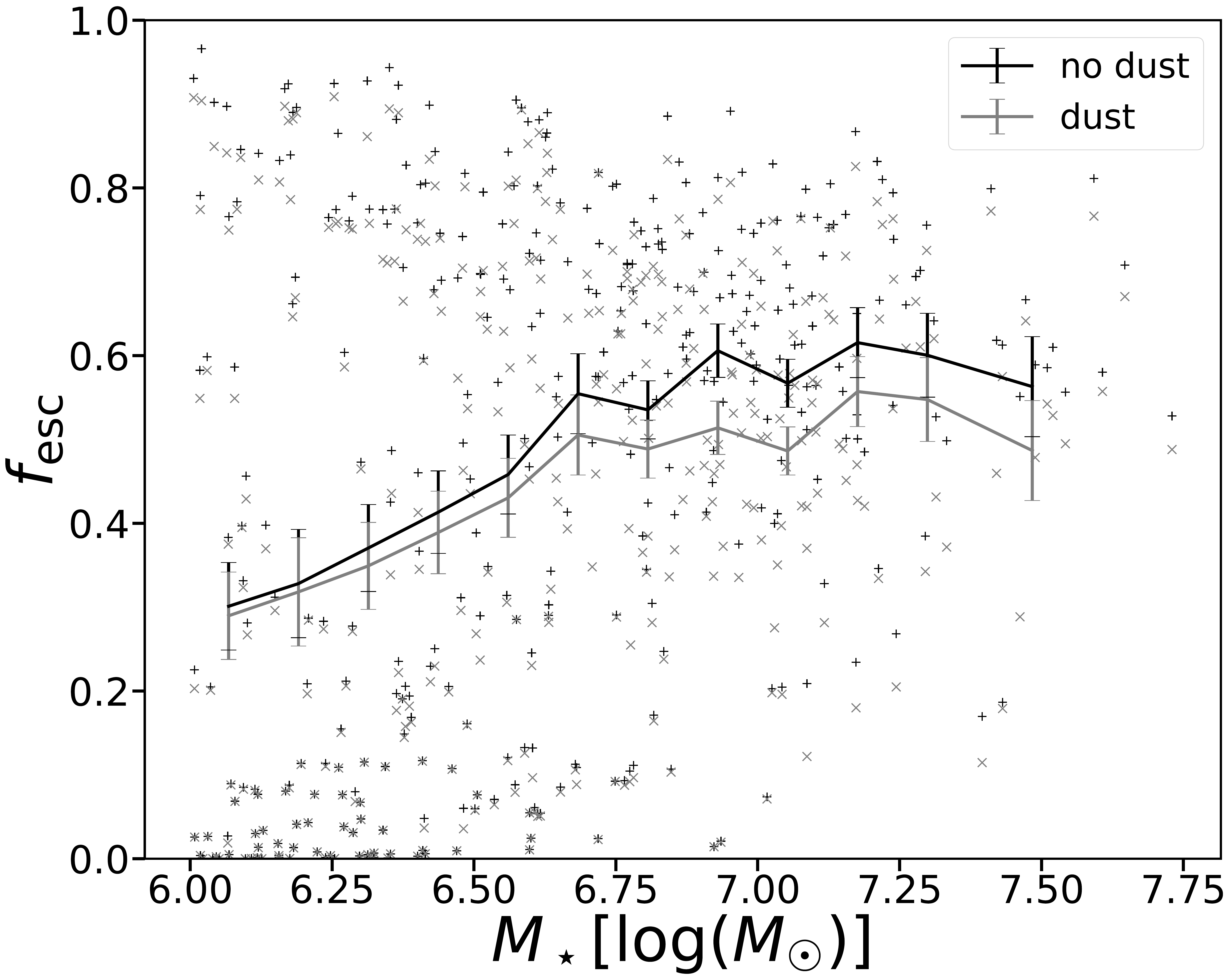} \vspace{0.5cm}
    \includegraphics[width=0.47\textwidth]{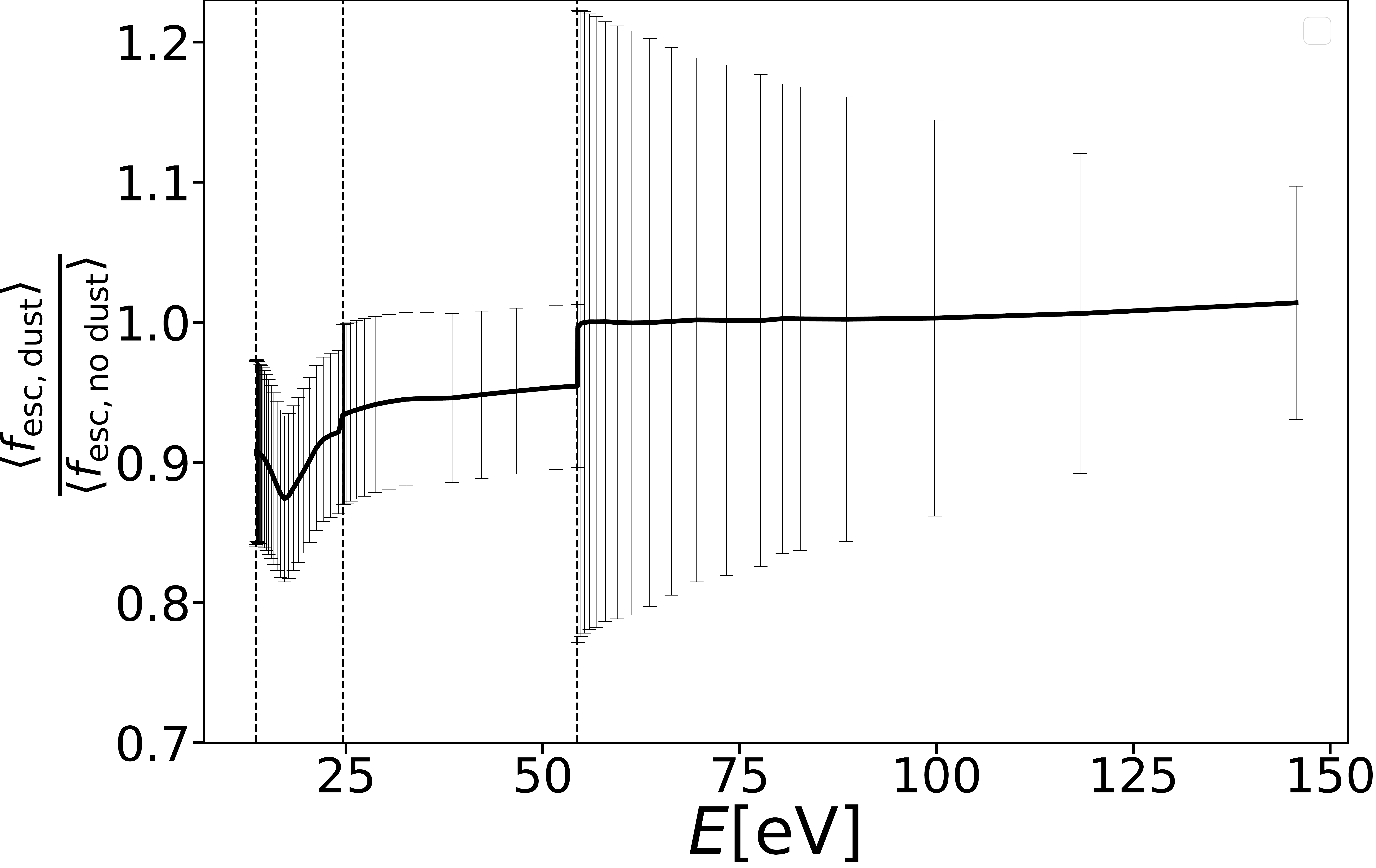}
    \caption{\textit{Top:} Halo escape fraction as a function of stellar mass with (grey) and without (black) dust absorption, as predicted by the TNG50+CRASH RT simulations. The scatter corresponds to individual halos at all redshifts, while the lines show the average escape fractions. Overall, dust reduces the escape fraction of ionizing radiation, although the magnitude of this effect is modest. \textit{Bottom:} Mean ratio of the halo escape fraction for dust versus no dust absorption, as a function of frequency. The error bars show the statistical uncertainty on the mean for each energy bin.}
    \label{fig:dust}
\end{figure}

The middle panel splits this comparison based on galaxy stellar mass, comparing $M_\star > 10^8$\,M$_\odot$ (blue lines) versus $M_\star < 5 \times 10^6$\,M$_\odot$ (red lines). The difference between emitted and escaping spectra at energies $E$ above 54.4eV is higher for the high-mass bin, than for the low-mass bin. As helium in all halos is mostly in the form of HeII rather than HeIII, photons with $E>54.4$eV in more massive halos typically travel longer distances, and hence encounter a larger optical depth, before reaching the virial radius than photons in smaller halos. Such an effect is much less visible at $E<54.4$eV because the gas is mainly in the form of HII and HeII and hence does not contribute to absorption in this range of energies. 

However, this scenario does not apply for halos with small escape fractions, where both hydrogen and helium are mainly in their neutral state. When we split this comparison based on halo escape fraction (bottom panel), we see that halos with $f_{\rm esc}<0.1$ have similar shapes in their emitted versus escaping spectra at all energies. In contrast, for high $f_\mathrm{esc}$ halos, we observe a clear difference at $E>54.4$eV.

Taken together, these results suggest the following scenario. Although the SED of stellar populations with binary stars has a significant tail of HeII ionizing photons which is absent when only single stars are considered \citep{gotberg20}, these photons are more easily absorbed in comparison to those with energies below $54.4\mathrm{eV}$. Thus, the larger fraction of HeIII found in simulations of reionization including binary stellar systems (e.g. \citealt{ma22}), might be reduced when accounting for a frequency dependent escape fraction. We note, though, that HeII reionization is driven by more energetic sources (i.e. mainly luminous black holes) and thus this caveat is not expected to have any major impact on the global reionization history.

\subsection{Impact of dust absorption}

Finally, we investigate the effect of dust on the escape fraction. To do so, we select a subset of halos and re-run the radiative transfer simulations including an additional dust component. The cross section of gas including dust is based on the model presented in \cite{glatzle21}, i.e. the Silicate-Graphite-PAH model \citep{weingartner01, draine07}. As total dust mass is not tracked in the TNG model, we adopt the simple prescription of a constant dust-to-metal (DTM) ratio of $40\%$, applied to each gas cell individually. In reality, the DTM ratio will be more complex, with dependencies on galaxy properties such as metallicity \citep{remy14}. Our model is a simple choice which enables to explore, and place a reasonable upper bound, on the contribution of dust to the absorption of ionizing radiation.

Figure \ref{fig:dust} shows the resulting trend of escape fraction versus stellar mass (top panel), contrasting the fiducial no dust case (gray) with the dust included model (black). At low stellar masses the escape fraction is only marginally affected by the presence of dust, with differences of order a few percent for $M_\star < 10^{6.5}$\,M$_{\astrosun}$. Additionally, we still observe the bimodality from Figure \ref{fig:med_fesc}, i.e. the existence of low-mass halos with very small escape fractions, further demonstrating that dust has little effect on $f_{\rm esc}$ in small halos.

However, dust becomes more important with increasing stellar mass. We observe a maximum impact on the escape fraction of $\sim\,10\%$, which occurs for galaxies with $M_{\star} = 10^7 $M$_{\astrosun}$ and remains fairly constant for even larger halos. This trend is consistent with higher mass halos having a higher total metal and thus dust content, such that the dust contribution to total absorption is larger in more massive halos \citep[see also][]{ma20}.

Figure \ref{fig:dust} also presents the impact of dust on the average halo escape fraction as a function of frequency (bottom panel). One can see that the strongest decrease in the halo escape fraction happens at energies between the HI and HeI ionization thresholds, where the presence of dust decreases the escape fraction by $10-15\%$. In particular, we observe a dip in the escape fraction with a minimum at $\approx 20$eV, roughly corresponding to the peak in the dust grain cross-section used in our model (see \citealt{glatzle21}). Between the HeII and HeIII ionization energies the suppression of the escape fraction remains roughly constant at $\approx 5-7\%$.

Finally, at energies higher than the HeII ionization threshold the contribution of dust to the total absorption becomes negligible. By comparing to Figure \ref{fig:freq_fesc} it is clear that this is due to the absorption of ionizing radiation being fully dominated by the abundant HeII which has a significantly larger contribution to the absorption of highly energetic ionizing radiation than dust \citep{glatzle21}.


\section{Summary and Conclusions} \label{sec:conlusions}

In this work we have investigated the dependence of the escape fraction of ionizing photons at halo scales, $f_{\rm esc}$, on different properties of the host galaxies and halos. To do this, we have post-processed halos extracted from the TNG50 simulation \citep{nelson19a,pillepich19} with the 3D multi-frequency radiative transfer code CRASH \citep{crash,crash2,crash3,glatzle21}. We have explored galaxies with stellar masses $10^6 \lesssim M_\star/{\rm M_\odot} \lesssim 10^{8.5}$ at redshifts $6 < z < 10$.

Our main results can be summarized as follows:

\begin{itemize}
    \item We find a large scatter of $f_{\rm esc}$, at fixed stellar and halo mass, across all redshifts and stellar masses in the range $10^6-10^8$M$_\odot$ studied herein. This scatter increases substantially for low mass galaxies ($M_\star \lesssim 10^7 \rm{M}_\odot$), where $f_{\rm esc}$ ranges from essentially zero to nearly 100\% across the galaxy population, due to the inherent stochasticity of the (young) stellar populations of these galaxies.
    \item The escape fraction rises with increasing galaxy (or halo) mass, from $\sim 20\%$ at $M_\star = 10^6$\,M$_\odot$ to a roughly constant value of $\sim 60\%$ at $M_\star = 10^7 - 10^8$\,M$_\odot$. Towards even higher masses, we see the hint of a turnover, after which escape fractions begin to decline.
    \item There is no consensus from cosmological (radiative/magneto)-hydrodynamical simulations for the value of $f_{\rm esc}$, nor its trend with galaxy mass. Our results are roughly consistent with some previous findings, and inconsistent with others.
    \item At fixed mass, the strongest predictor for the scatter in $f_{\rm esc}$ is the total ionizing emissivity of the halo. However, the escape fraction also depends on the details of the spatial distribution of (stellar) sources in relation to the underling gas distribution.
    \item Halos with a stellar (halo) mass $\lesssim 10^{7.5}$ ($10^{10}$) M$_\odot$ contribute the majority of ionizing photons at all redshifts. The global ionizing emissivity $\dot{N}_{\rm ion}$ obtained by summing the contributions of all halos can easily match observational constraints, for intermediate values of an unresolved local escape fraction $f_\mathrm{esc,loc}=0.3-0.7$.
    \item The absolute value of $f_{\rm esc}$ depends, non-linearly, on our adopted value for the local (unresolved) escape fraction, i.e. the attenuation of ionizing radiation below the resolution scale of the simulation.
    \item $f_{\rm esc}$ is strongly dependent on the photon energy. Due to the frequency dependence on the absorption cross-section and the abundant HeII present in most halos, its global average value is much smaller ($<\,$10-15\%) at $E>54.4$eV than at lower energies, where it is above $35\%$. This suggests that the impact on reionization from the high energy tail of binary stellar systems might be reduced when accounting for an energy dependent escape fraction.
    \item The presence of dust, when included self-consistently in our radiative transfer approach, reduces $f_{\rm esc}$. However, the magnitude of this effect is moderate, escape fractions decreasing by a few percent at $M_\star \lesssim 10^{6.5}$M$_\odot$, and up to 10\% for larger halos.
\end{itemize}

In this work we have opted to solve the radiative transfer in post-processing. While RHD simulations have the advantage of capturing the back-reaction of the radiation on the gas, they are extremely expensive in terms of computation time and thus generally employ a less accurate radiative transfer scheme (often applied only to the hydrogen component of the gas), as well as a fairly coarse resolution of the source spectrum (typically 1-3 frequency bins). Conversely, performing the radiative transfer in post-processing allows us to employ an accurate Monte Carlo algorithm to follow the propagation of ionizing photons, as well as employ a fine spectral resolution (64 bins), in addition to modeling dust and gas (in its hydrogen as well as helium components) absorption self-consistently. Due to the computational expense of our approach we have limited the analysis to halos with $M_{\rm vir}<10^{11}$M$_\odot$. However, we have seen in Figure \ref{fig:emisdensity_star} that such halos, because of their low abundance, do not provide a major contribution to the process of reionization. Hence, although intrinsically an interesting problem, precise knowledge of the escape fraction from high-mass halos is not critical for the modeling of cosmic reionization.

Currently, the main limitation in our modeling of the escape fraction is the treatment of the gas phase-structure in the dense interstellar medium. This arises from two related aspects: the numerical resolution of the TNG50 simulation, and the sub-grid model for the two-phase structure of star forming gas. As a result, our  findings depend on a local, or unresolved, escape fraction parameter. This choice applies to all simulation studies of escaping radiation. The absolute value of $f_\mathrm{esc}$ is ultimately dependent on the adopted prescription for the unresolved absorption. In this work we have presented the first systematic study of the effect of varying the sub-grid cloud-scale escape fraction on $f_{\rm esc}$, but more study is motivated. In the future, it will be imperative to better constrain and/or model $f_{\rm esc,loc}$ as a function of local gas properties, global galaxy or halo properties, and/or a combination thereof.

\section*{Data Availability}

Data directly related to this publication and its figures is available on request from the corresponding author. The IllustrisTNG simulations, including TNG50, are publicly available and accessible at \url{www.tng-project.org/data} \citep[see][]{nelson19a}.

\section*{Acknowledgements}

The authors are grateful to Enrico Garaldi for useful discussions, and also thank Aaron Smith for helpful comments.
DN acknowledges funding from the Deutsche Forschungsgemeinschaft (DFG) through an Emmy Noether Research Group (grant number NE 2441/1-1). This work is supported by the Deutsche Forschungsgemeinschaft (DFG, German Research Foundation) under Germany´s Excellence Strategy EXC 2181/1 - 390900948 (the Heidelberg STRUCTURES Excellence Cluster).
The TNG50 simulation was run with compute time granted by the Gauss Centre for Supercomputing (GCS) under Large-Scale Projects GCS-DWAR on the GCS share of the supercomputer Hazel Hen at the High Performance Computing Center Stuttgart (HLRS). GCS is the alliance of the three national supercomputing centres HLRS (Universit{\"a}t Stuttgart), JSC (Forschungszentrum J{\"u}lich), and LRZ (Bayerische Akademie der Wissenschaften), funded by the German Federal Ministry of Education and Research (BMBF) and the German State Ministries for Research of Baden-W{\"u}rttemberg (MWK), Bayern (StMWFK) and Nordrhein-Westfalen (MIWF). Additional simulations and analysis were carried out on the Freya machine of the Max Planck Institute for Astrophysics (MPA) and systems at the Max Planck Computing and Data Facility (MPCDF). 


\bibliographystyle{mnras}
\bibliography{refs}
\appendix
\section{Appendix}

\subsection{Numerical convergence}
\label{app:conv}

As discussed in Section \ref{sec:methods}, in the CRASH radiative transfer (RT) simulation, Monte Carlo sampled photon packets are propagated from stellar sources through the simulation domain. To obtain an accurate evaluation of gas properties, it is crucial that each voxel is crossed sufficiently often by a photon packet, i.e. the results are sensitive to the source sampling. To demonstrate that our results are numerically converged with respect to the photon packet number, we have thus run a subsample of halos with different numbers of photons packets emitted per source. 

\begin{figure}
    \centering
    \includegraphics[width=0.49\textwidth]{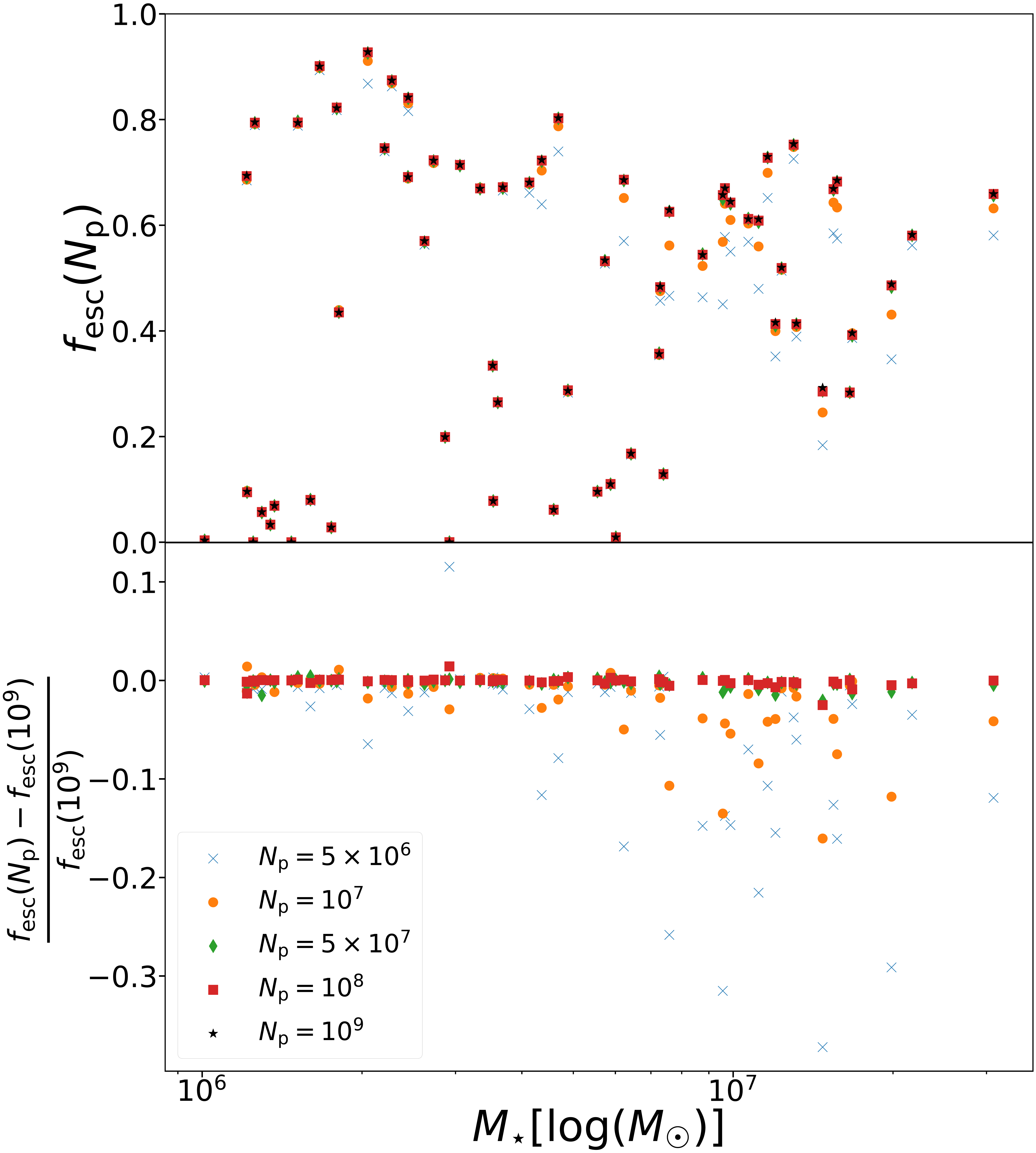}
    \caption{{\it Top panel:} Escape fraction as a function of stellar mass for different numbers of photon packets emitted per stellar particle. {\it Bottom panel:} Relative deviation of $f_\mathrm{esc}$ obtained with $N_p$ photons from $f_\mathrm{esc}$ obtained with $N_p=10^9$ photon packets.}
    \label{fig:photon_convergence}
\end{figure}

We show the resulting escape fractions in Figure \ref{fig:photon_convergence}, where we also plot the relative deviation of the escape fraction obtained with a given photon packet number from the highest setting, i.e. $N_\mathrm{p}=10^9$. We see that a strong convergence, with an error  $\lesssim 1\%$ for all mass bins, is reached only for $N_\mathrm{p} \gtrsim 5 \times 10^7$. To account for a possible increase in the error due to longer lines of sight in larger halos, we have therefore adopted a conservative $N_\mathrm{p}=10^8$ for our RT runs.

\subsection{Reduction of source numbers}

To limit the computational cost of the radiative transfer simulations, we have developed and tested an approach, similar to the one described in \cite{eide20}, which reduces the number of sources without affecting the final results. For each halo, all sources with an emissivity $Q_{\rm sources}$ less than a given percentage of the maximum source emissivity found within that halo are merged into nearby more luminous sources if they are within $N$ voxels, otherwise they are completely removed. We have run a number of tests to investigate the impact of a different choice for the merging distance.

\begin{figure}
    \centering
    \includegraphics[width=0.49\textwidth]{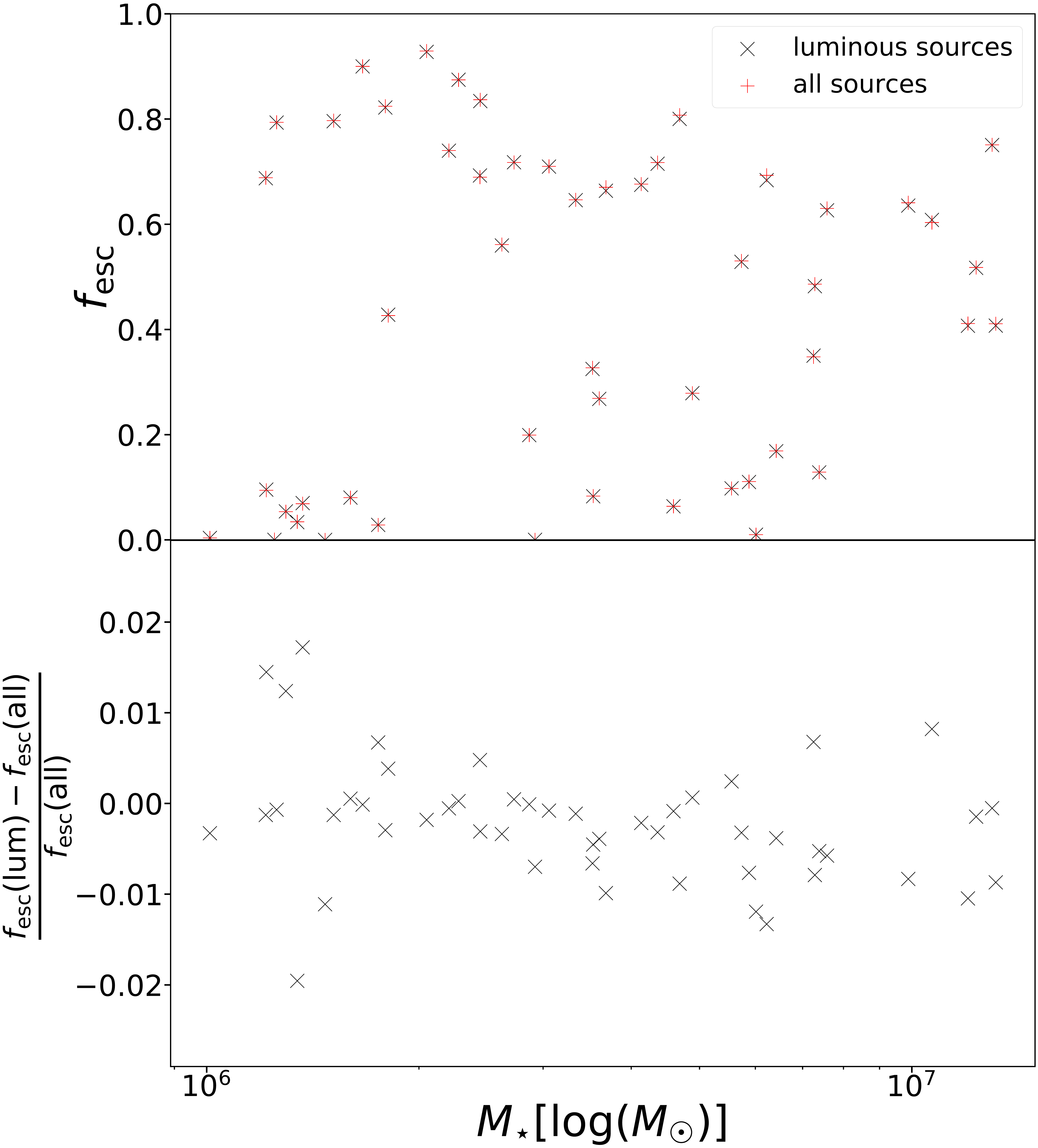}
    \caption{\textit{Top panel}: Escape fraction for a subsample of halos with source reduction as used in our production runs (black) and all the stellar particles in the simulation (red). \textit{Bottom panel}: Deviation from the escape fraction in the runs with reduced sources as compared to running the halos with all sources.}
    \label{fig:reduction_convergence}
\end{figure}

Figure \ref{fig:reduction_convergence} shows the effect of the source reduction on the escape fraction of a sample of halos for our reference case of a threshold percentage of 1\% and a merging distance of 5 voxels. We have verified that as long as the threshold percentage is below 1\%, more than 99\% of the halo emissivity is conserved, and our results are unchanged. 

\end{document}